\documentclass[twocolumn,showpacs,amsmath,prb]{revtex4-1}
\usepackage{graphicx}
\usepackage{subfigure}
\usepackage{epsfig}
\usepackage{dcolumn}
\usepackage{bm}
\usepackage{color}
\usepackage{xcolor}
\usepackage{hyperref}
\usepackage{txfonts}
\usepackage{physics}
\usepackage{siunitx}
\usepackage[titletoc]{appendix}
\usepackage{amsmath, amssymb, dsfont}

\allowdisplaybreaks

\definecolor{mygreen}{rgb}{0,0.5,0}
\definecolor{myblue}{rgb}{0,0,0.75}
\definecolor{mymagenta}{cmyk}{0,1,0,0.12}

\begin{document}

\title{Nonlinear one-way edge-mode interactions for frequency mixing in topological photonic crystals}
\author{Zhihao Lan}
\email{z.lan@ucl.ac.uk} \affiliation{Department of Electronic and Electrical Engineering,
University College London, Torrington Place, London, WC1E 7JE, United Kingdom}
\author{Jian Wei You}
\email{j.you@ucl.ac.uk} \affiliation{Department of Electronic and Electrical Engineering,
University College London, Torrington Place, London, WC1E 7JE, United Kingdom}
\author{Nicolae C. Panoiu}
\email{n.panoiu@ucl.ac.uk} \affiliation{Department of Electronic and Electrical Engineering,
University College London, Torrington Place, London, WC1E 7JE, United Kingdom}
\date{\today}
\begin{abstract}
Topological photonics aims to utilize topological photonic bands and corresponding edge
modes to implement robust light manipulation, which can be readily achieved in the linear regime of
light-matter interaction. Importantly, unlike solid state physics, the common test bed for new
ideas in topological physics, topological photonics provide an ideal platform to study wave mixing
and other nonlinear interactions. These are well-known topics in classical nonlinear optics but
largely unexplored in the context of topological photonics. Here, we investigate nonlinear
interactions of one-way edge-modes in frequency mixing processes in topological photonic crystals.
We present a detailed analysis of the band topology of two-dimensional photonic crystals with
hexagonal symmetry and demonstrate that nonlinear optical processes, such as second- and
third-harmonic generation can be conveniently implemented via one-way edge modes of this setup.
Moreover, we demonstrate that more exotic phenomena, such as slow-light enhancement of nonlinear
interactions and harmonic generation upon interaction of backward-propagating (left-handed) edge
modes can also be realized. Our work opens up new avenues towards topology-protected frequency
mixing processes in photonics.
\end{abstract}
\maketitle

\section{Introduction}

One of the most important developments in condensed matter physics in the past decades is the
discovery of topological insulating materials \cite{Kane_RMP10,Qi_RMP11}. These materials feature
gapped bulk but gapless edge modes, which propagate unidirectionally along the system edge and are
immune to local disorder, thus opening a promising avenue towards robust wave manipulation
protected by topology. Inspired by this development, the emerging field of topological photonics
aims to extend these topology related ideas to the realm of photonics
\cite{Lu_NP14,Shvets_NP17,Chen_PQE17, Gong_AOM17, Chen_OE18, Rider_JAP19, Ozawa_RMP19}, which holds
great promises for innovative optical devices by exploiting robust, scattering-free light
propagation and manipulation. As the concept of energy band exists at the single particle level
both in condensed matter physics and photonics, the goal of realizing photonic topological
insulators can be readily achieved in the linear regime of light matter interaction. Indeed,
topological phenomena of electromagnetic waves in a linear medium can be understood by mapping
Maxwell equations to the Schr\"{o}dinger equation \cite{Haldane_PRL08,Nittis_AP14}.
\begin{figure}[b!]
\centering
\includegraphics[width=\columnwidth]{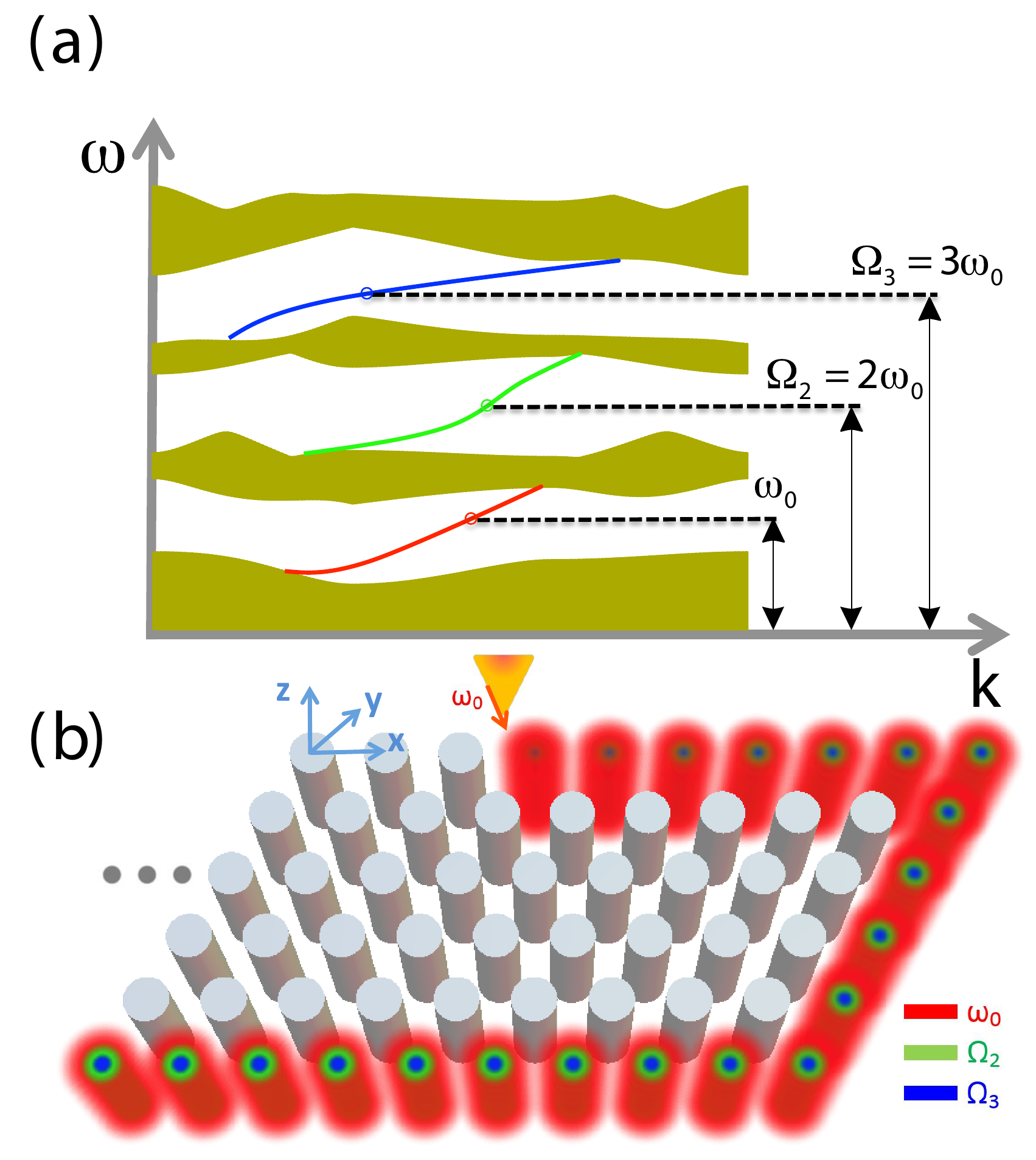}
\caption{(a) Schematic band structure showing the emergent edge modes due to the nontrivial
topology of the bulk frequency bands. The edge modes can couple \textit{via} SHG and THG frequency
mixing processes. (b) Real space illustration showing the unidirectional propagation of coupled
edge modes along the system edge. The red arrow indicates the excitation source of the fundamental
wave whereas the green and blue waves are generated as a result of nonlinear wave mixing.}
    \label{fig1}
\end{figure}

Photonics, however, has several features not present in solid-state physics. For example, optical
gain and loss can be utilized to implement non-Hermitian photonics based on parity-time symmetry
\cite{Ge_NP17}. The recently realized topological insulator laser demonstrates the power of this
new ingredient and could deepen our understanding of the interplay between non-hermiticity and
topology in active optical systems \cite{TIL_Science1,TIL_Science2}. Another well-known feature is
the existence of nonlinearity in many optical materials. In fact, optical nonlinear effects play a
key r\^{o}le in modern photonic applications, giving rise to a variety of important phenomena,
including the formation of solitons, modulation and all-optical switching of optical signals, and
frequency conversion for the generation of ultrashort pulses \cite{Boyd_Book08}. Thus one expects
new physics to emerge when adding nonlinearity to photonic systems with nontrivial topological
properties. Indeed, it has been shown that when a photonic topological insulator is embedded in an
optical medium with Kerr nonlinearity, lattice edge solitons could arise \cite{Lumer_PRL13,
Chong_PRL16}. The possibility to enhance the conversion efficiency of harmonic generation in the
presence of topological edge states has also been studied \cite{Qian_OE18,Chong_NC19,
yuri_natnano_19}. Moreover, traveling-wave amplifiers \cite{Peano_prx16}, topological sources of
quantum light \cite{Hafezi_nature18}, nonlinear control \cite{Yuri_PRL2018} and mapping
\cite{Yuri_PRL2019} of photonic topological edge states have also been achieved. Despite these
important advances, the feasibility of achieving nonlinear optical mixing of edge states of
topological photonic crystals via {\it phase matching}, which is one of the most fundamental
nonlinear optical processes, has not been explored yet. We would also like to highlight the key
differences between our work and previous works \cite{Qian_OE18, Chong_NC19, yuri_natnano_19,
Yuri_PRL2019} on nonlinear optics pertaining to topological edge states: the works \cite{Qian_OE18,
Chong_NC19, yuri_natnano_19} are based on one-dimensional (1D) systems, so that the topological
edge states are non-propagating optical modes localized at the edges of the 1D system, whereas the
system investigated in \cite{Yuri_PRL2019} is 2D, i.e., the same as ours, but there is only one
topological band gap at the fundamental frequency, and thus there are no nonlinear optical
interactions between topological modes.

In this work, we study nonlinear optical interactions of edge modes in topological photonic
crystals (PhCs), as per Fig.~\ref{fig1}. In particular, we present a detailed study of the band
topology of 2D photonic crystals with hexagonal symmetry by mapping out the Chern-number-graded gap
phase diagrams. Interestingly, we find that most gaps of the phase diagrams have exactly one edge
state in each gap, thus providing a convenient configuration to study the nonlinear interaction of
these modes. To this end, by properly tailoring the edge configuration to achieve phase matching,
we show that key nonlinear optical processes, such as second- and third-harmonic generation (SHG,
THG) can be readily realized in this setup. Beyond this proof-of-principle demonstration of these
nonlinear optical processes, we further show that some more exotic nonlinear optical phenomena can
also be observed in these topological PhCs, including slow-light enhanced frequency conversion
efficiency and higher-harmonic generation upon interaction of so-called backward-propagating
(left-handed) modes. All these novel ideas open up new avenues towards active photonic devices with
novel functionalities for photonic applications.

The article is organized as follows. In the next section we present and discuss the linear optical
properties of the topological photonic crystal, whereas in Sec.~\ref{nonlinprop} we describe the
nonlinear optical interaction between one-way topological modes and the coupled-mode theory that
governs the nonlinear propagation of interacting topological modes. Moreover, in Sec.~\ref{exp} we
briefly discuss possible experimental implementations of the ideas presented in this study, whereas
in the last section we summarize the main conclusions of this work.

\section{Linear optical properties of the investigated topological photonic crystal}\label{linprop}
In this section we describe the geometry and material parameters of the topological PhC
investigated in this work, as well as the topological properties of the bulk frequency bands and
edge topological modes.

\subsection {The system}

\begin{figure*}[t!]
\centering
\includegraphics[width=\textwidth]{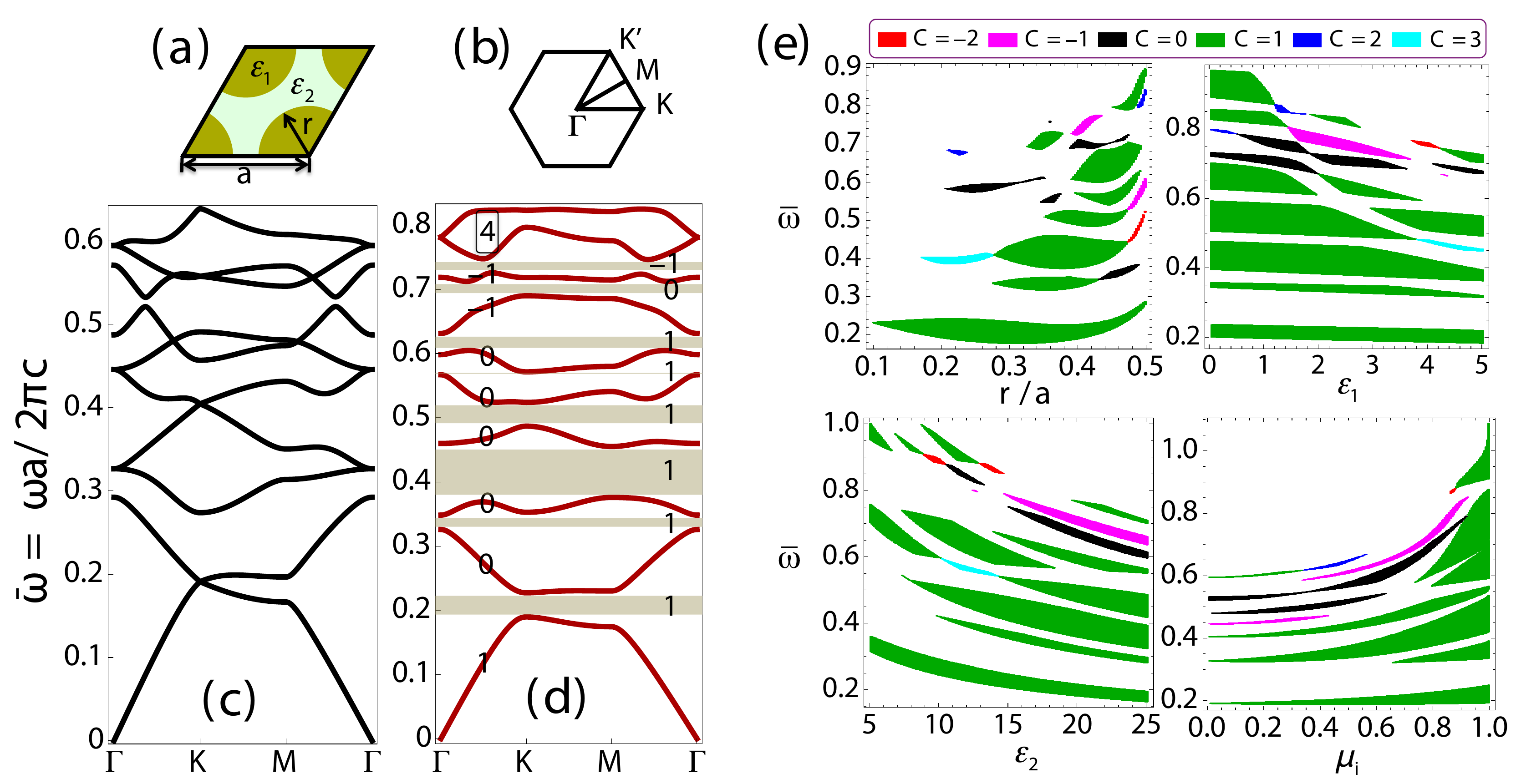}
\caption{(a) The unit cell with lattice constant $a$ of the PhC, where $r$ and $\epsilon_1$ are the
radius and relative permittivity of the cylinders, respectively, and $\epsilon_2$ and $\mu_i$ are
the relative permittivity and off-diagonal component of the relative permeability of the background
magnetic material, respectively. (b) The first Brillouin zone with the symmetry points, $\Gamma$,
$K$, $K^{\prime}$, and $M$. (c) Photonic band structure of the PhC, computed for $r=0.4a$,
$\epsilon_1=3$, $\epsilon_2=18$, and $\mu_i=0$. (d) Nontrivial topological bands for $\mu_i=0.8$
(the other parameters are the same as in (c)), where the Chern number of each band and the gap
Chern numbers are provided (except for the last two bands that touch each other and thus have the
same Chern number). (e) Chern-number-graded gap phase diagrams when varying $r$, $\epsilon_1$,
$\epsilon_2$, and $\mu_i$, determined for $r=0.4a$, $\epsilon_1=3$, $\epsilon_2=18$, and
$\mu_i=0.8$.} \label{fig2}
\end{figure*}

We begin by describing the system setup. First, 2D PhCs possessing topological frequency gaps
around frequencies of  $\omega_0$, $\Omega_2=2\omega_0$, and $\Omega_3=3\omega_0$ are designed in
order to study SHG and THG \textit{via} the corresponding edge modes located inside these gaps. As
such, in principle any PhC satisfying this condition could be employed. Nonetheless, it would be
beneficial if the first gap is topological since typically, the spectral separation among frequency
bands and the gap widths decrease as the frequency increases. In view of this, employing the
transverse magnetic (TM) modes of a PhC with hexagonal symmetry lattice, which features Dirac cones
at $K$ and $K^{\prime}$ points of the first Brillouin zone (FBZ) between the first and second bands
\cite{pc_book08}, is a natural choice. More specifically, one expects that for this configuration
the first gap becomes topological when gapping the Dirac cones by breaking the time-reversal
symmetry. Consequently, we consider triangular PhCs whose unit cell contains only one cylinder with
radius, $r$, as depicted in Fig.~\ref{fig2}a. Lattice structures with hexagonal symmetry but having
more cylinders in each unit cell, like honeycomb and Kagome lattices with two and three cylinders,
respectively, could potentially be employed, too. The second step of our design procedure is to
include magnetic and nonlinear materials. To guide potential experimental implementations and for
the sake of specificity, we consider cylinders with low-permittivity ($\epsilon_1$), non-magnetic
nonlinear material immersed in a magnetic background material with high-permittivity
($\epsilon_2$). Note that the permittivity of the cylinders has to be lower than that of the
background to ensure that Dirac cones exist.

\subsection{Topological properties of the bulk frequency bands}
We now move on to the topological properties of the bulk frequency bands of the proposed
non-magnetic ($\mu=\mu_{0}$) PhCs whose unit cell and FBZ are shown in Figs.~\ref{fig2}a and
\ref{fig2}b. In the following, we use normalized frequency and momentum, $\overline{\omega}=\omega
a/2\pi c$ and $\overline{k}=k a/\pi$, respectively, where $c$ is the speed of light and $a$ is the
lattice constant. Figure~\ref{fig2}c shows the photonic band structure of the PhC with $r=0.4a$,
$\epsilon_1=3$, and $\epsilon_2=18$ from which one can see the Dirac cone between the first and
second bands at $K$ and $K^{\prime}$ points. All band structures presented in this work were
calculated using COMSOL Multiphysics 5.3 \cite{COMSOL}, a commercial software package based on the
finite-element method, and validated using Synopsis's BandSOLVE software \cite{BandSOLVE}.

As known, the Chern number of each band is zero in systems with time-reversal symmetry
\cite{Wang_PRL08}. A common way to break time-reversal symmetry and generate bands with nonzero
Chern number is to use magnetic materials \cite{wang_Nature09, Chan_PRL11, Skirlo_PRL15, Bahari_Science17,
feifei_NC18}, where the permeability tensor of the material under an external magnetic field along
the $z$-axis possesses off-diagonal components in the $x-y$ plane, \textit{i.e.},
\begin{align}\label{mu}
    \mu=\left(
          \begin{array}{ccc}
            \mu_0 & i\mu_i & 0 \\
            -i\mu_i & \mu_0 & 0 \\
            0 & 0 & \mu_0 \\
          \end{array}
        \right).
\end{align}
Here, we set $\mu_0=1$ and take $\mu_i$ as a parameter to quantify the effect of time-reversal
symmetry breaking. Figure~\ref{fig2}d shows the photonic band structure for $\mu_i=0.8$, where one
can see that the Dirac cone is now gapped.

To characterize the topology of the frequency bands, we calculate the Chern number of the $n$th
band, defined as \cite{Haldane_PRL08,Lu_NP14}:
\begin{gather}\label{Cn}
C_n= \frac{1}{2\pi} \oiint_{FBZ} \mathcal{F}_n({\bf k}) \cdot d {\bf k},
\end{gather}
where $\mathcal{A}_n(\mathbf{k})=\expval{i\nabla_{\mathbf{k}}}{\mathbf{E}_n({\mathbf{k}})}$ and
$\mathcal{F}_n(\mathbf{k}) = \nabla_{\mathbf{k}} \times \mathcal{A}_n(\mathbf{k})$ are the Berry
connection and Berry curvature, respectively, with $\mathbf{E}_n({\mathbf{k}})$ being the electric
field of the $n$th band mode with momentum, $\mathbf{k}$. The momentum-space integral is performed
over the FBZ, whereas the inner product of the Berry connection is defined as
$\bra{\mathbf{E}_{\alpha}}\ket{\mathbf{E}_{\beta}}=\iint
\epsilon(\mathbf{r})\mathbf{E}_{\alpha}(\mathbf{r}) \cdot
\mathbf{E}_{\beta}(\mathbf{r})d\mathbf{r}$, with the real-space integral performed over the unit
cell. The Chern number is calculated using the algorithm described in \cite{Fukui_JPSJ05} (for more
details about this algorithm, see Appendix \ref{appsec:1}).
\begin{figure}
\centering
\includegraphics[width=\columnwidth]{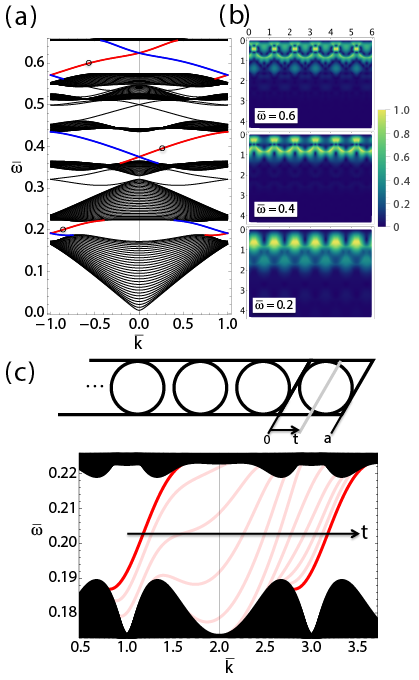}
\caption{(a) Photonic band structure of a 1D PhC strip that is periodic along the $x$-axis and has
finite size of 30 unit cells along the $y$-axis (top and bottom edges are terminated by a perfect
electric conductor at $r=0.42a$). The other simulation parameters are $\epsilon_1=3$,
$\epsilon_2=20$, and $\mu_i=0.8$. The edge modes in the three gaps around $\overline{\omega}=0.2$,
$0.4$, and $0.6$ are depicted by red and blue lines and are formed at the top and bottom edges of
the PhC, respectively. (b) Field profiles of the three one-way edge modes at
$\overline{\omega}=0.2$, $0.4$, and $0.6$ of the top edge. Exponential decay of the field around
the PhC edge can be observed (integers indicate the number of unit cells). (c) Dispersion curves of
edge modes can be tailored by changing the edge termination, as indicated in the sketch.}
\label{fig3}
\end{figure}

The calculated Chern numbers of the photonic bands are indicated in Fig.~\ref{fig2}d on top of each
band, and the gap Chern number, defined as the sum of the Chern numbers of the bands below the gap,
is also given for each gap. The gap Chern number characterizes the topology of the gap in the sense
that its sign determines the propagation direction of the edge modes and its value indicates the
number of edge states located inside the gap \cite{Lu_NP14}. An interesting feature revealed by
Fig.~\ref{fig2}d is that the first few gaps have Chern number $C=1$, which means that there is one
edge mode in each gap and all propagate in the same direction. Therefore, this configuration
provides a convenient platform to study nonlinear optical processes, \textit{e.g.}, SHG and THG.

To understand intuitively what regimes can be achieved with this setup, it is instructive to map
out the Chern-number-graded gap phase diagrams, defined as the variation of the gap Chern numbers
with the system parameters $r$, $\epsilon_1$, $\epsilon_2$, and $\mu_i$. We show these gap phase
diagrams in Fig.~\ref{fig2}e for $r=0.4a$, $\epsilon_1=3$, $\epsilon_2=18$, and $\mu_i=0.8$, when
one parameter is varied while keeping fixed the others. The results show that most domains of the
phase diagrams have $C=1$. Moreover, one can also see gaps with $C=2$, $C=3$ and, importantly, even
gaps with negative Chern numbers, $C=-1$ and $C=-2$. As we will demonstrate, this variety of values
of the gap Chern numbers leads to particularly rich physics when nonlinear interactions of
topological modes are considered.

\subsection{Topological properties of the edge modes}
Guided by the phase diagrams in Fig.~\ref{fig2}e, we choose the suitable parameters to create
photonic gaps suitable to study SHG and THG. According to the principle of bulk-edge correspondence
in systems with finite size, when the gap has nonzero Chern number, one-way edge modes will emerge
in the gap. We present in Fig.~\ref{fig3}a the photonic band structure of a PhC strip with 30 unit
cells along the $y$-axis and periodic along the $x$-axis. This figure illustrates the emergence of
various edge states across bulk photonic gaps in a range of frequencies. For the sake of clarity,
we mark in red and blue the edge states formed on the top and bottom edges of the PhC strip,
respectively. The field profiles of the edge states at frequencies $\overline{\omega}=0.2$,
$\overline{\Omega}_{2}=2\overline{\omega}=0.4$, and $\overline{\Omega}_{3}=3\overline{\omega}=0.6$,
presented in Fig.~\ref{fig3}b, highlight the key feature of the edge state -- exponential decay of
the field away from the edge.

A well-known prerequisite for achieving efficient frequency conversion processes is to phase match
the interacting waves \cite{Boyd_Book08}. In the current context, this requires a method to tune
the wave vectors of the edge modes. As far as we know this issue has not been previously discussed,
perhaps due to the irrelevance of phase matching in other (linear) physics involving edge modes. We
find that the wave vector of edge modes can be readily tuned by simply changing the configuration
of the edge termination. Figure~\ref{fig3}c shows how the edge-mode band of the top edge changes
when varying the location of the edge termination (see the sketch at the top of Fig.~\ref{fig3}c).
In general, we find that the edge-mode band shifts by about one reciprocal lattice vector,
$G=2\pi/a$, as one increases the width of the PhC strip by one unit cell. Note that due to the
periodicity of the system along the $x$-axis, one can always shift the wave vector of the edge mode
to the region of $[-\pi/a,~\pi/a]$ by adding a momentum of $nG$, with $n$ a suitable integer.

\section{Nonlinear optical interaction between one-way topological modes}\label{nonlinprop}
In this section we first introduce the coupled-mode theory (CMT) for SHG mediated by the
topological edge modes (a similar derivation for THG is given in the Appendix~\ref{appsec:2}). We
then present proof-of-concept results, both numerical and analytical, on topics, such as SHG and
THG upon edge-mode interaction, SHG in the slow-light regime, and SHG via interaction between
forward- and backward-propagating edge modes.

\subsection{Coupled-mode theory describing second-harmonic generation}\label{CMT-SHG}
The derivation of the CMT for SHG upon nonlinear interaction of topological edge-modes follows the
formulation of the CMT governing nonlinear pulse interactions in nonlinear PhC slab wavegudes
\cite{pmw10jstqe,lp16prb}. Thus, to begin with, we note that our system contains magneto-optic
materials, i.e., $\hat{\mu} \ne \hat{\mu}^T$, but rather $\hat{\mu}=\hat{\mu}^{\dagger}$, so that
we will use the conjugated form of the Lorentz reciprocity theorem \cite{LPT_book} for the vector
field $\mathbf{F}=\mathbf{E}_1^*\times \mathbf{H}_2+\mathbf{E}_2\times \mathbf{H}_1^*$, where
$\{\mathbf{E}_1,\mathbf{H}_1\}$ are the electric and magnetic fields of an eigenmode of the
topological PhC, whereas $\{\mathbf{E}_2, \mathbf{H}_2\}$ are the electric and magnetic fields
under the effect of nonlinear polarization $\mathbf{P}_{NL}$. Here, $\hat{\mu}$ is the magnetic
permeability tensor and the symbols ``$T$'' and ``$\dagger$'' indicate the transpose and Hermitian
transpose operations, respectively. Note also that as we consider 2D PhCs, we denote by
$\hat{\mathbf{z}}$ the out-of-plane direction and all the physical quantities only depend on the
$x$ and $y$ coordinates.

To start with, we write the eigenmodes of the topological PhC, both at the fundamental ($f$) and
second-harmonic ($s$) frequencies, denoted by $\omega_0$ and $\Omega_2=2\omega_0$, respectively, in
their Bloch forms, i.e.,
\begin{subequations}\label{FFfields}
\begin{align}
\mathbf{E}_1^f(\mathbf{r},\omega_0)&=\frac{\mathbf{e}_{f}(\mathbf{r},\omega_0)}{\sqrt{P_f}}e^{i k_{f}(\omega_0) x},\\
\mathbf{H}_1^f(\mathbf{r},\omega_0)&=\frac{\mathbf{h}_{f}(\mathbf{r},\omega_0)}{\sqrt{P_f}}e^{i
k_{f}(\omega_0) x},
\end{align}
\end{subequations}
and
\begin{subequations}\label{SHfields}
\begin{align}
\mathbf{E}_1^s(\mathbf{r},\Omega_2)&=\frac{\mathbf{e}_{s}(\mathbf{r},\Omega_2)}{\sqrt{P_s}}e^{i k_{s}(\Omega_2) x},\\
\mathbf{H}_1^s(\mathbf{r},\Omega_2)&=\frac{\mathbf{h}_{s}(\mathbf{r},\Omega_2)}{\sqrt{P_s}}e^{i
k_{s}(\Omega_2) x}.
\end{align}
\end{subequations}
In these equations, $\{\mathbf{e}_{f}, \mathbf{h}_{f}\}$ and $\{\mathbf{e}_{s}, \mathbf{h}_{s}\}$
are the lattice-periodic electric and magnetic Bloch fields of the fundamental and second-harmonic
waves, respectively, $k_{f,s}$ are the corresponding Bloch wavevectors, assuming that the waves
propagate along the $x$-axis, and the vector $\mathbf{r}$ lies in the transverse $(x,y)$-plane. If
we choose the normalization constants $P_{f,s}$ such that
\begin{equation}
\frac{1}{4}\int_{-\infty}^{\infty} \left(\mathbf{e}_{f,s}^*\times \mathbf{h}_{f,s}+\mathbf{e}_{f,s}
\times \mathbf{h}_{f,s}^* \right)  \cdot \hat{\mathbf{x}}dy =P_{f,s},
\end{equation}
the modes $\{\mathbf{E}_1^f,\mathbf{H}_1^f\}$ and $\{\mathbf{E}_1^s,\mathbf{H}_1^s\}$ in
Eqs.~\eqref{FFfields} and \eqref{SHfields}, respectively, carry \SI{1}{\watt} per unit length along
the longitudinal $z$-axis.

The mode power is related to the mode energy per unit length contained in one unit cell, $W$, and
the group velocity, $v_g$, via the relation:
\begin{equation}\label{power}
P_{f,s}=\frac{W_{f,s}}{a}v_g^{f,s}=\frac{W_{f,s}^{\textrm{el}}+W_{f,s}^{\textrm{mag}}}{a}v_g^{f,s},
\end{equation}
where, for non-dispersive media, the electric and magnetic energies are given by the following
formulae:
\begin{subequations}\label{emenergy}
\begin{align}
W_{f,s}^{\textrm{el}}&=\frac{1}{4}\int_{A_{cell}}\mathbf{e}_{f,s}^{*}\cdot\hat{\epsilon}\mathbf{e}_{f,s}dA, \\
W_{f,s}^{\textrm{mag}}&=\frac{1}{4}\int_{A_{cell}}\mathbf{h}_{f,s}^{*}\cdot\hat{\mu}
\mathbf{h}_{f,s}dA.
\end{align}
\end{subequations}

One can also define an effective width of the mode, $w^{\textrm{eff}}$, in terms of the Poynting
vector of the field \cite{lp16prb},
\begin{equation}
w^{\textrm{eff}}_{f,s}=\frac{{\left(\int_{-\infty}^{\infty}\vert \mathbf{e}_{f,s} \times
\mathbf{h}_{f,s} \vert dy\right)}^2}{\int_{-\infty}^{\infty}\vert \mathbf{e}_{f,s} \times
\mathbf{h}_{f,s} \vert^{2}dy}.
\end{equation}

From the Maxwell equations, one can readily infer that, in the frequency domain, the optical modes
satisfy the following equations:
\begin{eqnarray}
\nabla \times \mathbf{E}_1=i\omega \hat{\mu}\mathbf{H}_1, \\
\nabla \times \mathbf{H}_1=-i\omega \hat{\epsilon}\mathbf{E}_1,
\end{eqnarray}
where $\omega=\omega_0$ and $\omega=\Omega_2$ for the fundamental and second-harmonic waves,
respectively.

Following the CMT, for the perturbed problem, we write the electric and magnetic fields of the
fundamental and second-harmonic waves as:
\begin{subequations}\label{CMTFFfields}
\begin{align}
\mathbf{E}_2^f(\mathbf{r},\omega_0)&=A_f(x)\frac{\mathbf{e}_{f}(\mathbf{r},\omega_0)}{\sqrt{P_f}}e^{ik_{f}(\omega_0)x}, \\
\mathbf{H}_2^f(\mathbf{r},\omega_0)&=A_f(x)\frac{\mathbf{h}_{f}(\mathbf{r},\omega_0)}{\sqrt{P_f}}e^{ik_{f}(\omega_0)x},
\end{align}
\end{subequations}
and
\begin{subequations}\label{CMTSHfields}
\begin{align}
\mathbf{E}_2^s(\mathbf{r},\Omega_2)&=A_s(x) \frac{\mathbf{e}_{s}(\mathbf{r},\Omega_2)}{\sqrt{P_s}}e^{ik_{s}(\Omega_2)x}, \\
\mathbf{H}_2^s(\mathbf{r},\Omega_2)&=A_s(x)
\frac{\mathbf{h}_{s}(\mathbf{r},\Omega_2)}{\sqrt{P_s}}e^{ik_{s}(\Omega_2)x},
\end{align}
\end{subequations}
where $A_f(x)$ and $A_s(x)$ are the slowly-varying envelopes of the fundamental and second-harmonic
waves, respectively, under the effect of  the nonlinear polarization $\mathbf{P}_{NL}$. The power
per unit length carried by the fundamental and second-harmonic fields are $\vert A_f(x)\vert^{2}$
and $\vert A_s(x)\vert^{2}$, respectively, meaning that these field amplitudes are measured in
$\sqrt{\mathrm{W/m}}$.

The Maxwell equations for the perturbed fields are:
\begin{eqnarray}
\nabla \times \mathbf{E}_2=i\omega \hat{\mu}\mathbf{H}_2, \\
\nabla \times \mathbf{H}_2=-i\omega \hat{\epsilon}\mathbf{E}_2-i\omega
\mathbf{P}_{NL}^{\omega},
\end{eqnarray}
where $\omega=\omega_0$ and $\omega=\Omega_2$ for the fundamental and second-harmonic waves,
respectively. Moreover, starting from the equation for the second-order nonlinear polarization in
the time domain,
$\mathbf{P}_{NL}(\mathbf{r},t)=\hat{\chi}^{(2)}(\mathbf{r},t):\mathbf{E}(\mathbf{r},t)\mathbf{E}(\mathbf{r},t)$,
where $\hat{\chi}^{(2)}$ is the second-order susceptibility tensor, one can easily show that the
nonlinear polarization at the fundamental and second-harmonic frequencies can be written as:
\begin{subequations}\label{CMTpol}
\begin{align}
\mathbf{P}_{NL}^{f}(\mathbf{r},\omega_0)&=\frac{2A_{f}^{*}(x)A_s(x)}{\sqrt{P_f P_s}}\hat{\chi}^{(2)}:\mathbf{e}_{f}^{*}(\mathbf{r},\omega_0) \mathbf{e}_s(\mathbf{r},\Omega_2) e^{i(k_s-k_f)x},\\
\mathbf{P}_{NL}^{s}(\mathbf{r},\Omega_2)&=\frac{A_f^{2}(x)}{P_f}\hat{\chi}^{(2)}:\mathbf{e}_f(\mathbf{r},\omega_0)
\mathbf{e}_f(\mathbf{r},\omega_0) e^{2ik_f x}.
\end{align}
\end{subequations}

We now use the 2D form of the divergence theorem, which states that for any general function $\bf{F}$,
\begin{equation}\label{dt}
\int_A \nabla\cdot \mathbf{F} dA=\frac{\partial}{\partial x} \int_A \mathbf{F}\cdot\hat{\mathbf{x}}
dA + \oint_{\partial A}\mathbf{F}\cdot\hat{\mathbf{n}} dl,
\end{equation}
where $A$ is an arbitrary cross-section perpendicular to the direction of wave propagation,
$\hat{\mathbf{x}}$, and $\hat{\mathbf{n}}$ is the unit vector outwardly normal onto $\partial A$ in
the plane of $A$. If we take $A$ to extend to infinity along the $y$-axis, the line integral
vanishes for fields that decay exponentially to infinity. Moreover, the l.h.s. of Eq.~\eqref{dt}
can be written as:
\begin{equation}\label{divth}
\int_{-\infty}^{\infty} \nabla\cdot \mathbf{F} dy = i\omega\int_{-\infty}^{\infty}
\mathbf{E}_{1}^{*}\cdot\mathbf{P}_{NL}^{\omega}dy.
\end{equation}

In deriving this equation, we used the vector identity $\nabla\cdot (\mathbf{A} \times \mathbf{B})=
\mathbf{B} \cdot (\nabla \times \mathbf{A})-\mathbf{A} \cdot (\nabla \times \mathbf{B})$ and the
fact that, since  $\hat{\epsilon}$ and $\hat{\mu}$ are Hermitian, the identities
$\mathbf{E}_{1}^{*}\cdot\hat{\epsilon}\mathbf{E}_2=(\hat{\epsilon}\mathbf{E}_1)^*\cdot\mathbf{E}_2$
and $\mathbf{H}_{1}^{*}\cdot\hat{\mu}\mathbf{H}_2=(\hat{\mu}\mathbf{H}_1)^*\cdot\mathbf{H}_2$ hold.

Let us now consider Eq.~\eqref{divth}, written for the fundamental frequency:
\begin{equation}\label{f1}
\int_{-\infty}^{\infty} \nabla\cdot \mathbf{F} dy = \frac{2i\omega_0A_{f}^{*} A_s
e^{i\Delta kx}}{P_f\sqrt{P_s}}\int_{-\infty}^{\infty} \mathbf{e}_{f}^{*}\cdot
\hat{\chi}^{(2)}(\omega_0;-\omega_0,\Omega_2):\mathbf{e}_{f}^{*} \mathbf{e}_s dy,
\end{equation}
where $\Delta k=k_{s}(\Omega_2)-2k_{f}(\omega_0)$ is the wavevector mismatch. Moreover, the r.h.s.
of Eq.~\eqref{dt} can be cast as:
\begin{align}\label{f2}
\frac{\partial}{\partial x}& \int_{-\infty}^{\infty} \mathbf{F}\cdot\hat{\mathbf{x}}dy =
\frac{\partial}{\partial x} \int_{-\infty}^{\infty} \left(\mathbf{E}_1^{f*}\times
\mathbf{H}_2^f+\mathbf{E}_2^f\times \mathbf{H}_1^{f*}\right) \cdot\hat{\mathbf{x}}dy \nonumber \\=
&\frac{dA_f(x)}{dx}\frac{1}{P_f}\int_{-\infty}^{\infty}\left(\mathbf{e}_{f}^{*}\times \mathbf{h}_f
+\mathbf{e}_f\times \mathbf{h}_{f}^{*}\right)\cdot\hat{\mathbf{x}}dy=4\frac{dA_f(x)}{dx}.
\end{align}
Comparing Eqs.~\eqref{f1} and \eqref{f2}, we arrive at the equation describing the slowly-varying
mode amplitude $A_f(x)$:
\begin{equation}\label{shg11}
 \frac{d A_f(x)}{dx} =i\gamma^{(2)}_{f}(x)A_{f}^{*}(x)A_s(x)e^{i\Delta kx},
\end{equation}
where the nonlinear coefficient at the fundamental frequency is
\begin{equation}\label{shg12}
\gamma^{(2)}_{f}(x)=\frac{\omega_0}{2P_f\sqrt{P_s}}\int_{-\infty}^{\infty}
\mathbf{e}_{f}^{*}\cdot \hat{\chi}^{(2)}(\omega_0;-\omega_0,\Omega_2):\mathbf{e}_{f}^{*}
\mathbf{e}_s dy.
\end{equation}

The equation governing the evolution of the slowly-varying mode amplitude of the second-harmonic,
$A_s(x)$, is derived in a similar way. Thus, when $\omega=\Omega_{2}$, the l.h.s. of Eq.~\eqref{dt}
becomes:
\begin{equation}\label{f3}
\int_{-\infty}^{\infty}\nabla\cdot\mathbf{F}dy = \frac{i\Omega_2 A_{f}^{2}e^{-i\Delta
kx}}{P_f\sqrt{P_s}}\int_{-\infty}^{\infty} \mathbf{e}_{s}^{*}\cdot
\hat{\chi}^{(2)}(\Omega_2;\omega_0,\omega_0):\mathbf{e}_{f}\mathbf{e}_f dy,
\end{equation}
and the r.h.s. of Eq.~\eqref{dt} can be expressed as:
\begin{align}\label{f4}
\frac{\partial}{\partial x} \int_{-\infty}^{\infty} &\mathbf{F}\cdot\hat{\mathbf{x}}dy =
\frac{\partial}{\partial x} \int_{-\infty}^{\infty} \left(\mathbf{E}_1^{s*}\times
\mathbf{H}_2^s+\mathbf{E}_2^s\times \mathbf{H}_1^{s*}\right) \cdot\hat{\mathbf{x}}dy \nonumber \\=
&\frac{dA_s(x)}{dx}\frac{1}{P_s}\int_{-\infty}^{\infty}\left(\mathbf{e}_{s}^{*}\times \mathbf{h}_s
+\mathbf{e}_s\times \mathbf{h}_{s}^{*}\right)\cdot\hat{\mathbf{x}}dy=4\frac{dA_s(x)}{dx}.
\end{align}

Finally, from Eqs.~\eqref{f3} and \eqref{f4}, we obtain the governing equation for the
slowly-varying amplitude $A_s(x)$:
\begin{equation}\label{shg21}
\frac{d A_s(x)}{dx} =i\gamma^{(2)}_s(x)A_{f}^{2}(x)e^{-i\Delta kx},
\end{equation}
where the nonlinear coefficient at the second-harmonic is
\begin{equation}\label{shg22}
\gamma^{(2)}_s(x)=\frac{\Omega_2}{4 P_f\sqrt{P_s}} \int_{-\infty}^{\infty}
\mathbf{e}_{s}^{*}\cdot
\hat{\chi}^{(2)}(\Omega_2;\omega_0,\omega_0):\mathbf{e}_{f}\mathbf{e}_f dy.
\end{equation}

In order to better understand the effects of the slow light on the strength of the nonlinear
interaction, we use Eq.~\eqref{power} in conjunction with Eq.~\eqref{emenergy} to express the
nonlinear coefficients as follows:
\begin{subequations}\label{nonlcoeff2}
\begin{align}
\gamma^{(2)}_{f}(x)&=\frac{4Z_{0}^{3/2}\omega_0n_{g,f}\sqrt{n_{g,s}}}{\sqrt{a}}\chi_{\mathrm{eff},f}^{(2)}(x)=\frac{4\omega_0}{\sqrt{a}\epsilon_{0}^{3/2}v_{g,f}\sqrt{v_{g,s}}}\chi_{\mathrm{eff},f}^{(2)}(x),\\
\gamma^{(2)}_{s}(x)&=\frac{2Z_{0}^{3/2}\Omega_2n_{g,f}\sqrt{n_{g,s}}}{\sqrt{a}}\chi_{\mathrm{eff},s}^{(2)}(x)=\frac{2\Omega_2}{\sqrt{a}\epsilon_{0}^{3/2}v_{g,f}\sqrt{v_{g,s}}}\chi_{\mathrm{eff},s}^{(2)}(x),
\end{align}
\end{subequations}
where $Z_{0}$ is the vacuum impedance, $n_{g,f/s}=c/v_{g,f/s}$ are the group indices of the two
interacting modes, and the effective second-order susceptibilities
$\chi_{\mathrm{eff},f/s}^{(2)}(x)$ are defined by the following relations:
\begin{widetext}
\begin{subequations}\label{effchi2}
\begin{align}
\label{effchi2f}\chi_{\mathrm{eff},f}^{(2)}(x)&=\frac{\displaystyle a^{2}\int_{-\infty}^{\infty}
\mathbf{e}_{f}^{*}\cdot \hat{\chi}^{(2)}(\omega_0;-\omega_0,\Omega_2):\mathbf{e}_{f}^{*}
\mathbf{e}_s dy}{\displaystyle \int_{A_{cell}}\left(\mathbf{e}_{f}^{*}\cdot\hat{\epsilon}_{r}\mathbf{e}_{f}+Z_{0}^{2}\mathbf{h}_{f}^{*}\cdot\hat{\mu}_{r}\mathbf{h}_{f}\right)dA {\left[\int_{A_{cell}}\left(\mathbf{e}_{s}^{*}\cdot\hat{\epsilon}_{r}\mathbf{e}_{s}+Z_{0}^{2}\mathbf{h}_{s}^{*}\cdot\hat{\mu}_{r}\mathbf{h}_{s}\right)dA\right]}^{1/2}},\\
\label{effchi2s}\chi_{\mathrm{eff},s}^{(2)}(x)&=\frac{\displaystyle a^{2}\int_{-\infty}^{\infty}
\mathbf{e}_{s}^{*}\cdot \hat{\chi}^{(2)}(\Omega_2;\omega_0,\omega_0):\mathbf{e}_{f}\mathbf{e}_f
dy}{\displaystyle
\int_{A_{cell}}\left(\mathbf{e}_{f}^{*}\cdot\hat{\epsilon}_{r}\mathbf{e}_{f}+Z_{0}^{2}\mathbf{h}_{f}^{*}\cdot\hat{\mu}_{r}\mathbf{h}_{f}\right)dA
{\left[\int_{A_{cell}}\left(\mathbf{e}_{s}^{*}\cdot\hat{\epsilon}_{r}\mathbf{e}_{s}+Z_{0}^{2}\mathbf{h}_{s}^{*}\cdot\hat{\mu}_{r}\mathbf{h}_{s}\right)dA\right]}^{1/2}}.
\end{align}
\end{subequations}
\end{widetext}

It should be noted that the nonlinear coefficients and nonlinear effective susceptibilities vary
with the $x$-coordinate over a characteristic length equal to the lattice constant, $a$, whereas
the characteristic length over which the field amplitudes vary is equal to $1/\Delta k$. When the
two interacting waves are nearly phase-matched, $a\ll 1/\Delta k$. Therefore, in order to describe
the nonlinear mode interactions, it is convenient to introduce the averaged physical quantities
\begin{subequations}\label{average}
\begin{align}
\overline{\gamma}^{(2)}_{f,s}&=\frac{1}{a}\int_{x_{0}}^{x_{0}+a}\gamma^{(2)}_{f,s}(x)dx,\\
\overline{\chi}^{(2)}_{\mathrm{eff},f/s}&=\frac{1}{a}\int_{x_{0}}^{x_{0}+a}\chi_{\mathrm{eff},f/s}^{(2)}(x)dx,
\end{align}
\end{subequations}
where $x_{0}$ is arbitrary. Then, by averaging Eqs.~\eqref{shg11} and \eqref{shg21}, we arrive at
the system of CMT governing the nonlinear dynamics of the interacting modes:
\begin{subequations}\label{CMTaverage2}
\begin{align}
\frac{d \overline{A}_f(x)}{dx}&=i\overline{\gamma}^{(2)}_{f}\overline{A}_{f}^{*}(x)\overline{A}_s(x)e^{i\Delta kx},\\
\frac{d \overline{A}_s(x)}{dx}&=i\overline{\gamma}^{(2)}_s\overline{A}_{f}^{2}(x)e^{-i\Delta kx},
\end{align}
\end{subequations}
where $\overline{A}_{f}(x)$ and $\overline{A}_{s}(x)$ are the averaged mode amplitudes at the
fundamental and second-harmonic frequencies, respectively.
\begin{figure*}
\centering
\includegraphics[width=\textwidth]{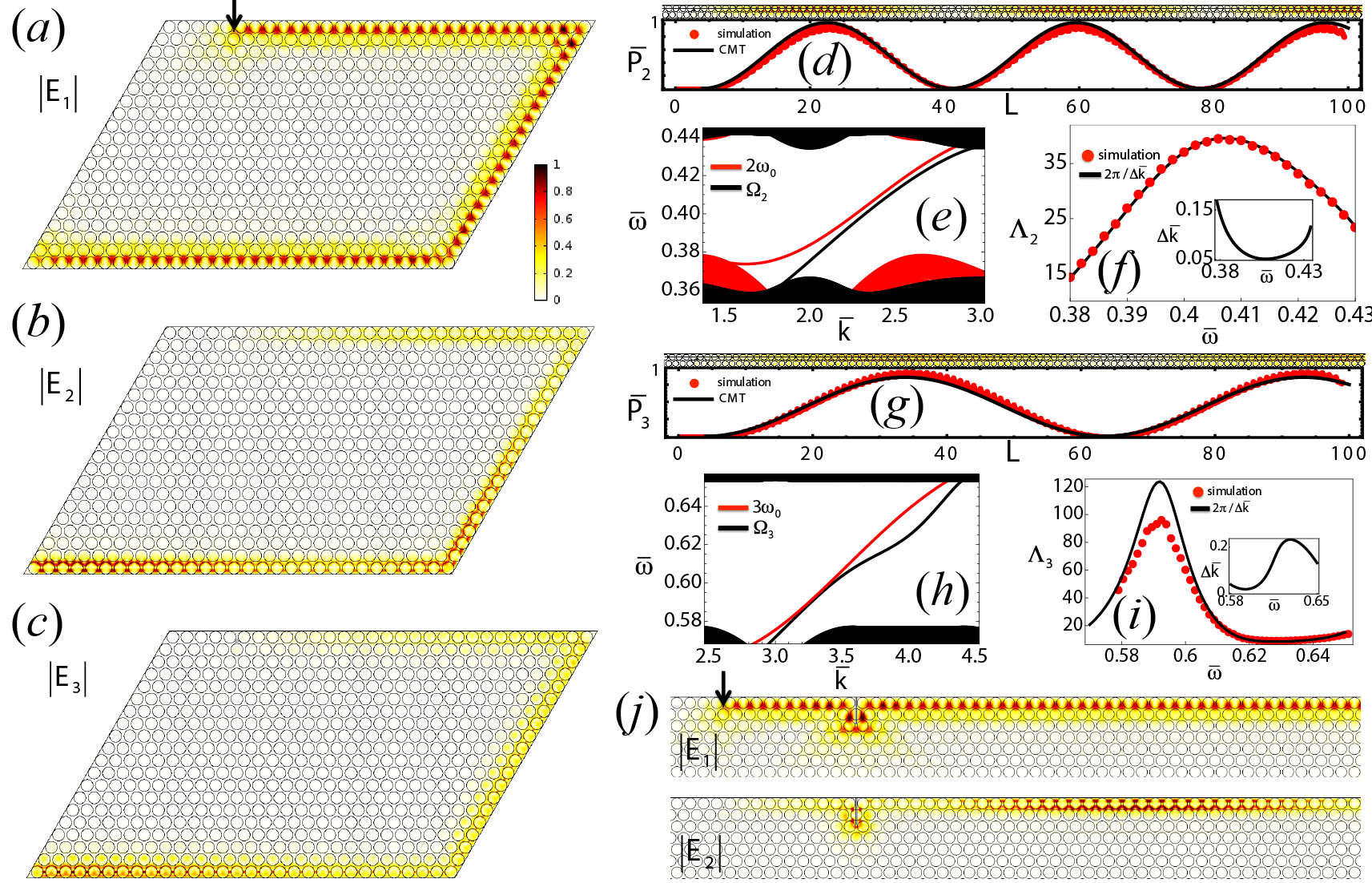}
\caption{(a-c) Simulated field profile intensities of $\mathbf{E}_1$, $\mathbf{E}_2$, and
$\mathbf{E}_3$ at $\overline{\omega}=0.2$, $\overline{\Omega}_{2}=2\overline{\omega}=0.4$, and
$\overline{\Omega}_{3}=3\overline{\omega}=0.6$, respectively, where $\mathbf{E}_1$ is induced by an
external source indicated by an arrow, whereas $\mathbf{E}_2$ and $\mathbf{E}_3$ are generated by
the corresponding nonlinear polarizations. The SHG and THG are simulated separately with
$\chi^{(2)}=\SI{e-21}{\coulomb\per\square\volt}$ and
$\chi^{(3)}=\SI{e-30}{\coulomb\meter\per\cubic\volt}$. In the simulations, absorbing boundary
condition (ABC) is used for the left edge and PEC for the other edges. (d-f) Detailed analysis of
SHG, where (d) shows the field intensity profile and power $\overline{P}_2$ (normalized by its peak
value) at SH in a much larger simulation domain so as to resolve the oscillations of
$\overline{P}_2$ due to a small phase mismatch $\Delta \overline{k}$ of the edge modes at FF
$\overline{\omega}=0.2$ and SH $\overline{\Omega}_{2}=0.4$ (dotted and solid lines correspond to
full numerical simulations and CMT, respectively)); (e) shows the edge modes at FF $\omega_0$
(plotted in terms of $2\omega_0$) and for the SHG, $\Omega_2$; (f) shows the effect of phase
matching, where the theoretically calculated $\Lambda_2=2\pi/\Delta \overline{k}$ and numerically
extracted oscillation period of $\mathbf{E}_2$ are in excellent agreement. The
$\overline{\omega}(\Delta \overline{k})$ curve is calculated according to the edge modes presented
in (e). (g-i) The results corresponding to (d-f), respectively, but calculated for the THG. (j)
Illustration of the effect of a structural defect (simulated as thin PEC rectangle) on the SHG
interaction of topologically protected modes. Comparing to the defect-free case shown in (d), one
can see that the oscillation period of the amplitude of the second-harmonic mode after bypassing
the defect is the same as that in (d).} \label{fig4}
\end{figure*}

\subsection{Description of the full-wave numerical simulations approach}
The full-wave dynamics of the nonlinear interaction of topological edge-modes (SHG and THG) were
determined numerically using the module  ``\textsf{Electromagnetic Waves, Frequency Domain}" of
COMSOL Multiphysics. Thus, to simulate the nonlinear frequency mixing processes in COMSOL, we
defined two ``\textsf{Electromagnetic Waves, Frequency Domain}" models, one for the fundamental
frequency $\omega_0$ and one for the second (third) harmonic frequency $\Omega_2$ ($\Omega_3$). The
two models are coupled using a ``\textsf{Polarization}" feature added to each of the models. We
assumed that for both the SHG and THG cases the nonlinear susceptibilities are diagonal tensors,
the diagonal elements being $\chi_2$ and $\chi_3$, respectively.

For the study of the SHG (where we consider TM-polarized modes), the nonlinear polarizations at the
FF and SH are:
\begin{subequations}\label{Pnonl2}
\begin{align}
P_{\mathrm{NL},z}^{\omega_{0}}&=2\chi_2 E_{2z} E_{1z}^*, \\
P_{\mathrm{NL},z}^{\Omega_{2}}&=\chi_2 E_{1z}^{2}.
\end{align}
\end{subequations}

For the study of THG, the corresponding nonlinear polarizations are:
\begin{subequations}\label{Pnonl3}
\begin{align}
P_{\mathrm{NL},z}^{\omega_{0}}&=3\chi_3 E_{3z} {E_{1z}^*}^{2},\\
P_{\mathrm{NL},z}^{\Omega_{3}}&=\chi_3 E_{1z}^{3}.
\end{align}
\end{subequations}

\subsection{Second-harmonic generation and third-harmonic generation upon edge-mode interaction}
We now investigate nonlinear frequency conversion processes \textit{via} the edge modes indicated
in Fig.~\ref{fig3}a using full-wave numerical simulations of Maxwell equations (for simulation
details, see Appendix \ref{appsec:1}), with the results being summarized in Fig.~\ref{fig4}. In the
following, we mainly focus on the discussion of SHG, as the results of THG can be understood
similarly.

We consider cylinders made of homogeneous and isotropic nonlinear material with a scalar nonlinear
second-order susceptibility of $\chi^{(2)}=\SI{e-21}{\coulomb\per\square\volt}$ (typical value of
$\chi^{(2)}$ varies from \SIrange{e-24}{e-21}{\coulomb\per\square\volt} \cite{Saleh_Book07}). The
pump electric field $\mathbf{E}_1$ is induced by an external source whereas $\mathbf{E}_2$ by the
nonlinear polarization at the SH, generated by $\mathbf{E}_1$. The amplitude of $\mathbf{E}_1$ is
chosen such that the undepleted pump approximation holds, \textit{i.e.} the amplitude of
$\mathbf{E}_1$ is much larger than that of $\mathbf{E}_2$ and thus $\mathbf{E}_1$ is roughly
constant during the frequency conversion process. Note, however, that our analysis remains valid
when this condition is not fulfilled, too, our choice being chiefly guided by a more facile
comparison between numerical and theoretical results, which is possible in this propagation regime.

From Figs.~\ref{fig4}a and \ref{fig4}b, one can observe that the field profiles of $\mathbf{E}_1$
and $\mathbf{E}_2$ are indeed the same as the profiles of the edge modes shown in Fig.~\ref{fig3}b,
indicating that the two edge modes are indeed nonlinearly interacting \textit{via} the SHG -- a key
result of our work. The physics of this nonlinear process can be accurately captured by the CMT
(see Appendix \ref{appsec:3}). In particular, the period of spatial oscillations of the SH field
$\mathbf{E}_2$ is determined by the wave-vector mismatch $\Delta k=k_{SH}-2k_{FF}$ ($\Delta
k=k_{TH}-3k_{FF}$ for THG). As a result, we can straightforwardly compare the numerically extracted
oscillation period of $\mathbf{E}_2$ with the theoretical prediction of $\Lambda_2=2\pi/\Delta
\overline{k}$, thus confirming that the key physics of nonlinear frequency conversion processes is
validated by our simulations.

We further validate these conclusions using a much larger simulation domain, with the corresponding
results being presented in Figs.~\ref{fig4}d--\ref{fig4}f. In Fig.~\ref{fig4}d, where
$\overline{\omega}_0=0.2$, $\overline{\Omega}_2=0.4$, and $\Delta \overline{k}=0.054$, we have
$\Lambda_2=37$. The agreement between the predictions the CMT and direct numerical simulations, of
both the period and amplitude of SH power oscillation along the propagation distance, is excellent.
We calculate $\Delta \overline{k}$ for all the frequencies of the interacting edge modes from
Fig.~\ref{fig4}e and present the theoretically calculated and numerically extracted oscillation
period $\Lambda_2$ in Fig.~\ref{fig4}f. An excellent agreement between the two results can be
observed, which confirms the key physics of SHG, namely phase matching is indeed at work in our
photonic system and SHG purely \textit{via} nonlinear interaction of edge modes occurs in our
setup. We also confirm the edge mode mediated THG as shown in Figs.~\ref{fig4}g--\ref{fig4}i, where
the discrepancy in Fig.~\ref{fig4}i between the numerical and theoretical results is due to
inherent limitations of numerical simulations at very small $\Delta k$.

While the SHG and THG of one-way edge modes are governed by the usual mechanism of topological
protection from the chiral nature of the edge modes, where the modes can bypass structural defects,
such as the sharp bends shown in Figs.~\ref{fig4}a-c, without undergoing backscattering, the effect
of structural defects on the coherence properties of nonlinear processes is an important but less
understood problem. In order to answer this key question, and taking SHG as an example, we
introduce a structural defect at a certain location along the edge where the fundamental and
second-harmonic modes co-propagate and present the corresponding simulation results in
Fig.~\ref{fig4}j. Thus, it can be seen that, as expected, both the fundamental and second-harmonic
modes bypass the defect without experiencing backscattering. More importantly, by comparing the
results in Figs. \ref{fig4}j and \ref{fig4}d, one can infer that the coherence length of the SHG,
namely the oscillation period of the amplitude of the interacting modes, is not altered by the
interaction with the structural defect. In other words, the coherence of the nonlinear interaction
is preserved in the presence of defects, meaning that the phase-matched nature of the nonlinear
mode interaction process is immediately regenerated after the mode interaction with each defect. As
the coherence length crucially determines the conversion efficiency of the nonlinear processes,
topological protection of the coherence length in nonlinear frequency mixing processes demonstrates
a new area where topology can boost the performance of photonic devices based on nonlinear optical
processes.
\begin{figure}
\centering
\includegraphics[width=\columnwidth]{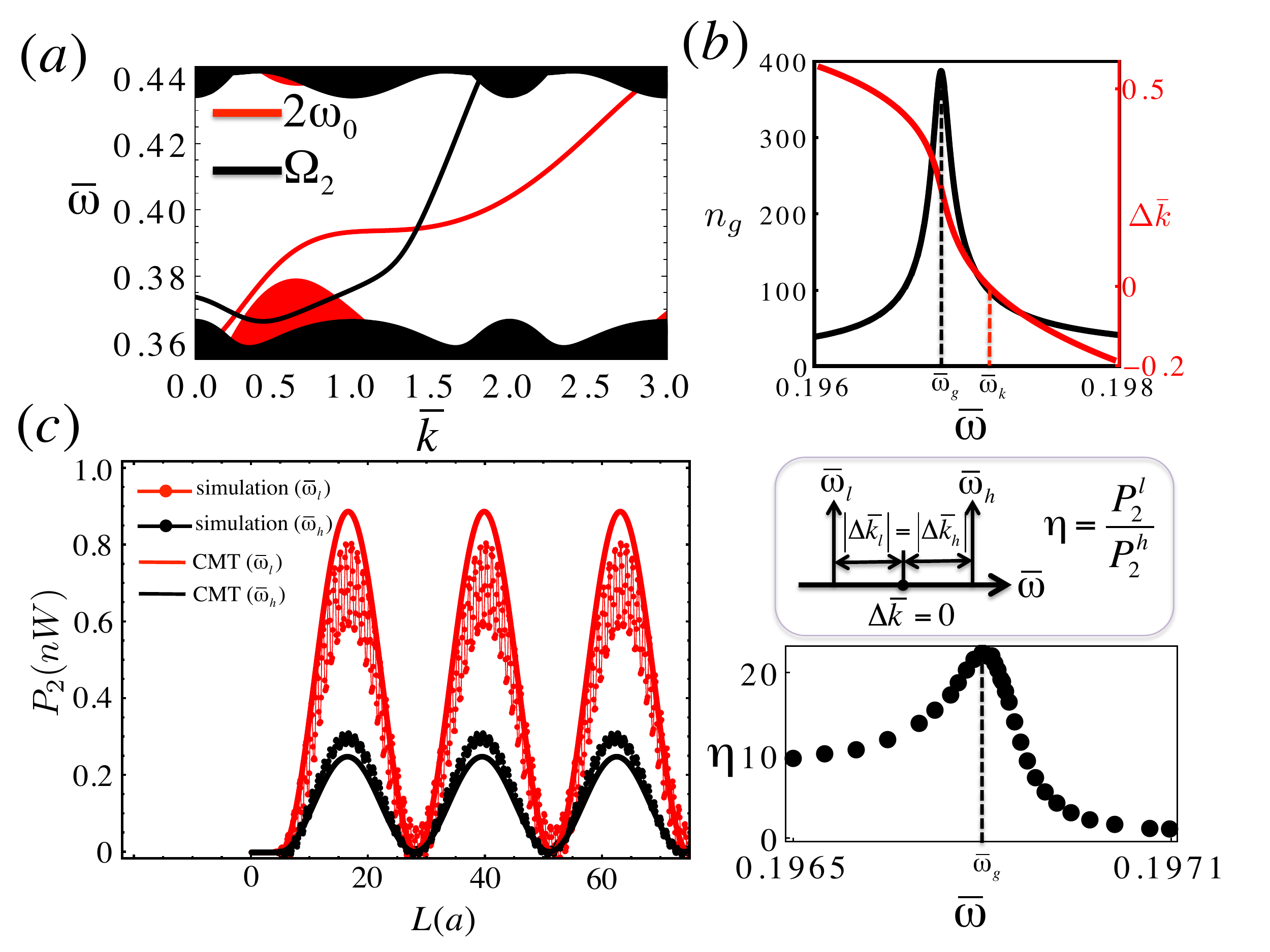}
\caption{(a) Dispersion of edge modes at $\omega_0$ and $\Omega_2$ determined for $t=0.24a$, as in
Fig.~\ref{fig3}c, whereas the other simulation parameters are the same as those in Fig.~\ref{fig4}.
(b) The dispersion curve of the FF mode now shows a plateau leading to a peak of $n_g$ at
$\overline{\omega}_g=0.1968$. The frequency of the phase-matching point
$\overline{\omega}_k\equiv\overline{\omega}(\Delta \overline{k}=0)$ is
$\overline{\omega}_k=0.1972$. (c) The enhancement of the SH conversion efficiency due to the
slow-light effect. The left panel shows an example where the SH power $P_2$ at two different
frequencies, $\overline{\omega}_l=0.1960$ and $\overline{\omega}_h=0.1975$, shows the same
oscillation period, yet its oscillation amplitude at $\overline{\omega}_l$ (closer to the maximum
of $n_g$, located at $\overline{\omega}_g$) is enhanced compared to that at $\overline{\omega}_h$.
The right panels show schematically the formal definition of the enhancement factor $\eta$ (top)
and its frequency dependence in the slow-light regime (bottom).}\label{fig5}
\end{figure}

Another relevant question, which we do not intend to fully answer here, is how other types of
perturbations affect the phase-matching condition of the nonlinear optical interaction of
topological modes. From a practical point of view, the most relevant such perturbation is
structural disorder, and in this context we consider two cases, namely weak and strong disorder.
The case of weak disorder can be analyzed using perturbative methods. Thus, let us assume that in
the absence of disorder the frequency-dependent wave-vector mismatch, $\Delta k_{i}(\omega)$,
vanishes at a certain frequency $\omega_{0}$, that is $\Delta k_{i}(\omega_{0})=0$. Then, adding
weak disorder to the unperturbed system will change the modal dispersion of the interacting modes,
such that the wave-vector mismatch varies, say, by a small quantity $\delta k(\omega)$.
Furthermore, since one expects that $\delta k(\omega)$ has constant sign around the frequency
$\omega_{0}$, it can be seen that there is a frequency $\omega_{0}+\delta\omega_{0}$ at which the
wave-vector mismatch in the presence of disorder, $\Delta k_{f}(\omega)=\Delta k_{i}(\omega)+\delta
k(\omega)$, vanishes, that is $\Delta k_{f}(\omega_{0}+\delta\omega_{0})=0$. To validate this
argument, one would have to employ full-wave simulations of the nonlinear optical interaction of
topological modes and average the results over an ensemble of disorder realizations large enough to
achieve convergence of the ensemble average. This computational analysis would be rather unfeasible
as just a single full-wave simulation requires several days to complete. On the other hand, for
large values of disorder strength, concepts such as optical modes, topological properties, and
photonic band gaps cease to exist, and therefore we do not investigate this case any further.

\subsection{Second-harmonic generation in the slow-light regime}
The slow-light regime, characterized by a significantly reduced group velocity, $v_g=d\omega/dk$,
can be particularly effective in enhancing the efficiency of nonlinear wave interactions. In the
context of SHG, this can be achieved when $v_{g}$ is reduced at one or both interacting waves. As
Fig.~\ref{fig3}c suggests, when one varies the location of the edge termination, the shift of the
dispersion curve of the edge modes is accompanied  by a change of its shape. For example, we find
that for $t=0.24a$, the dispersion curve of the edge mode at the FF has a plateau (see
Fig.~\ref{fig5}a) leading to a peak of the group index, $n_g=c/v_g$, at $\overline{\omega}_g$ (see
Fig.~\ref{fig5}b, where we define the slow-light regime by the condition $n_g>20$ \cite{lp16prb}).
On the other hand, the fact that the phase-matching spectral point
$\overline{\omega}_k\equiv\overline{\omega}(\Delta \overline{k}=0)\neq\overline{\omega}_g$ provides
a valuable approach to study the interplay between slow-light and phase-matching effects in the
enhancement of SH conversion efficiency.

To this end, we fix the power of FF at $P_1=\SI{1}{\watt}$ and compare the conversion efficiencies
at two frequencies $\overline{\omega}_l$ and $\overline{\omega}_h$, lower and higher as compared to
$\overline{\omega}_k$, respectively, chosen such that $\Delta
\overline{k}(\overline{\omega}_l)=-\Delta \overline{k}(\overline{\omega}_h)$ (see the sketch in
Fig.~\ref{fig5}c). As such, phase-matching has the same influence on the wave interaction in the
two cases and the conversion efficiency enhancement,
$\eta=P_2(\overline{\omega}_l)/P_2(\overline{\omega}_h)$, is purely due to slow-light effects. The
results in Fig.~\ref{fig5}c show that slow-light contribution to $\eta$ can be larger than
$20\times$. Alternative scenarios to enhance the SH conversion efficiency can also be devised,
e.g., by using PhCs for which $\overline{\omega}_k=\overline{\omega}_g$ or PhCs for which the edge
modes at both the FF and SH are slow-light edge modes.
\begin{figure}[b!]
\centering
\includegraphics[width=\columnwidth]{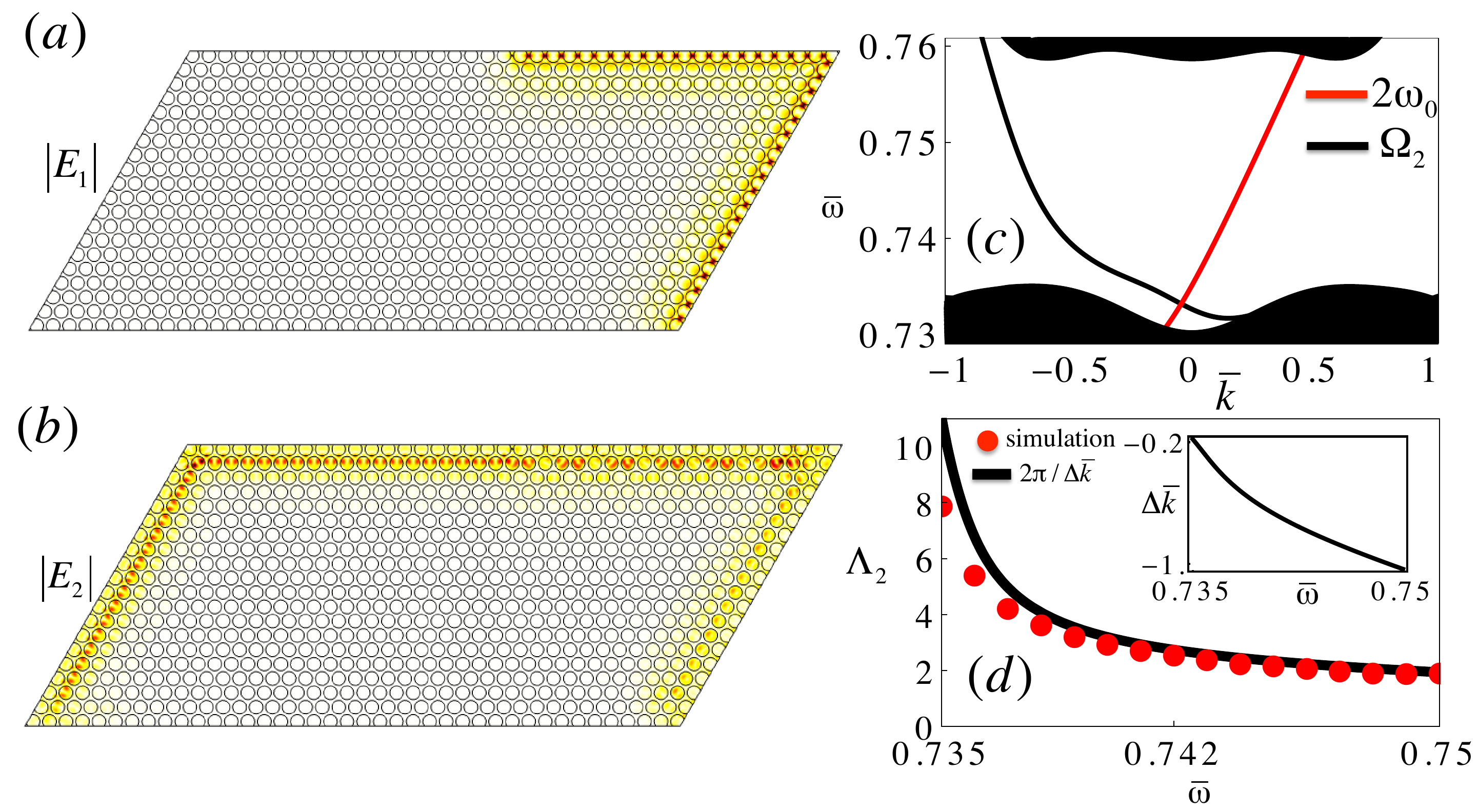}
\caption{SHG \textit{via} interaction between forward- and backward-propagating edge modes. This
setting exploits the existence of a gap with negative Chern number at the SH for $r=0.41a$,
$\epsilon_1=3$, $\epsilon_2=20$, $\mu_i=0.82$ (see Fig.~\ref{fig2}e) and edge termination at
$t=0.82a$ (see Fig.~\ref{fig3}c). (a-b) Field intensity profiles of $\mathbf{E}_1$ and
$\mathbf{E}_2$ calculated for $2\overline{\omega}_0=0.74$, illustrating that whereas the mode at
the FF propagates clockwise ($C=1$ for the $\omega_0$ gap), the edge mode at the SH is
backward-propagating ($C=-1$ for the $\Omega_2$ gap). In the simulation, ABC is used for the bottom
edge while PEC boundaries for the other edges. (c) Edge states of the two gaps, which show the
hallmark of edge modes with negative Chern number: the slope of the dispersion curve of the edge
mode at the SH is negative. (d) Comparison between the theoretically calculated and the numerically
extracted oscillation period $\Lambda_2$, confirming that the phase matching mechanism is involved
in this unusual nonlinear interaction regime.} \label{fig6}
\end{figure}

\subsection{Second-harmonic generation \textit{via} interaction between forward- and backward-propagating edge modes}
We now move on to an important class of nonlinear processes, which are challenging to achieve in
regular optical media, and demonstrate SHG \textit{via} interaction of backward-propagating edge
modes. To this end, we exploit the existence of photonic gaps with negative Chern number in our
system, as per Fig.~\ref{fig2}e. In particular, we explore a case where the gap at the FF has
$C=1$, while the gap at the SH has $C=-1$. The corresponding edge modes are shown in
Fig.~\ref{fig6}c, which illustrates the signature of modes with negative Chern number,
\textit{i.e.}, the slope of the mode dispersion curve (at $\Omega_2$) is negative. The simulation
results of the field profiles are presented in Figs.~\ref{fig6}a and \ref{fig6}b. It clearly shows
that whereas the mode at the FF propagates clockwise, as it is a forward-propagating mode, the SH
wave propagates counterclockwise because in the left-half region of the simulation domain there is
no field at the FF and consequently no nonlinear polarization at the SH (note that we placed the
source of the FF wave at the middle of the top edge and used absorbing boundary condition for the
bottom edge). We also confirm that the phase matching mechanism is involved in this unusual mode
interaction regime, as per the results in Fig.~\ref{fig6}d. This backward-propagating mode regime
is promising for practical applications where one requires to separate the FF mode from the mode
generated at the SH.

\section{Experimental considerations}\label{exp}
The key idea of this work involves coupling the edge states within different topological frequency
band gaps via optical nonlinearity. This central idea of our study is rather general and can
potentially be implemented in different experimental platforms available for topological photonics.
In this section, for the sake of completeness, we first present a set of materials and parameters
that can be used to experimentally implement our model system. Then we discuss further possible
experimental platforms that can be used to implement the theoretic ideas and results presented in
our work.

Since we separate the magnetic and nonlinear material components as the background and cylindrical
regions of the PhC, respectively, possible experimental implementations could be readily conceived
considering the fact that at microwave frequencies, magnetic materials to realize topological band
gaps \cite{wang_Nature09, Chan_PRL11, Skirlo_PRL15, Bahari_Science17, feifei_NC18} and nonlinear
materials to realize frequency mixing \cite{Boyd_PRL71, Bridges_APL72, Boyd_APL72, Ahn_JAP83} are
routinely used. For example, to demonstrate the topology-protected SHG, one can use as the
background medium yttrium iron garnet (YIG) \cite{microwave_book}, a magnetic material widely used
in recent experiments in topological photonics \cite{wang_Nature09, Chan_PRL11, Skirlo_PRL15,
Bahari_Science17, feifei_NC18}. This material has relative permittivity $\epsilon=15$ and
saturation magnetization of $4\pi M_{s}=\SI{1780}{Gauss}$. At frequency of \SI{10}{\giga\hertz} and
an external magnetic field of $H_0=\SI{1000}{Oe}$ along the $z$-axis, the components of the
permeability of YIG in the $x-y$ plane are $\mu_{\text{diag}}=0.85$ and
$\mu_{\text{off-diag}}=0.54$. For the nonlinear material, one could use NaNO$_2$, which has
relative permittivity of $\epsilon=4.18$ and $\chi^{(2)}=\SI{3.2e-22}{\coulomb\per\square\volt}$ at
frequency of tens of \si{\giga\hertz} \cite{Boyd_PRL71}. Setting the radius of the cylinders to
$r=0.35a$, with $a\approx\SI{3}{\milli\meter}$, the system has topological band gaps at
$\overline{\omega}_0\in [0.195,~0.23]$ and $\overline{\Omega}_2 \in [0.385,~0.41]$, so that SHG is
achieved in the frequency interval of $\overline{\omega}_0\in [0.195,~0.205]$. The effects of
material losses and frequency dispersion in a ferrite at microwave frequencies have been discussed
in Ref.~\cite{Wang_PRL08}, showing that for YIG, the decay length is around $1300a$, thus far
exceeding practical structural dimensions, and the band gap width slightly decreases by about
\SI{6}{\percent}.

Furthermore, recent advances in different experimental platforms for topological photonics
\cite{Lu_NP14,Shvets_NP17,Chen_PQE17, Gong_AOM17, Chen_OE18, Rider_JAP19, Ozawa_RMP19} provide a
variety of choices available to implement the idea of topologically protected nonlinear frequency
mixing processes in diverse photonic systems. More specifically, as nonlinearity exists in many
optical materials and is easy to incorporate in an experimental platform, the key task reduces to
designing photonic systems with several topological frequency band gaps that can satisfy the
frequency and phase matching conditions. Regarding this requirement, we stress that several recent
experiments have demonstrated the existence of topological band gaps using coupled waveguides or
resonators \cite{Rechtsman_Nature13, Hafezi_NP13, Noh_PRL18}. In these tight-binding systems,
further topological band gaps at higher frequencies can readily be created by considering waveguide
or resonance modes at higher frequencies.

In what follows, we outline and briefly discuss several such experimental configurations in which
this can potentially be achieved. \textit{i) Nonlinear and magnetic metamaterials}: magnetism and
nonlinearity can be easily implemented using metamaterials \cite{Smith_PRL2000, Linden_Science06,
Lapine_RMP13, Gao_PRL15}, which provides an alternative to using regular materials, such as YIG and
NaNO$_2$ suggested above. For example, to create nonlinear metamaterials, approaches, such as
insertion of nonlinear elements, nonlinear host medium, and nonlinear transmission lines are
commonly used \cite{Lapine_RMP13}. Artificial magnetism at optical frequencies can be created using
metamaterials based on split-ring resonators \cite{Smith_PRL2000}. Harmonic generation and
topological edge states have also been studied in such types of metamaterials
\cite{Linden_Science06, Gao_PRL15}. \textit{ii) Graphene plasmonic crystals} \cite{Yeung_NanoL14,
Jin_PRL17, Pan_NC17, You_arxiv19}: in Ref.~\cite{Jin_PRL17} (see Fig.~3c therein) it has been
demonstrated the existence of topological edge modes within different band gaps of a graphene
plasmonic crystal. Moreover, in another study \cite{Yeung_NanoL14} (see Fig.~2a therein) the
existence of two Dirac points at the frequencies of \SI{3.4}{\tera\hertz} and \SI{6.8}{\tera\hertz}
has been demonstrated. These Dirac points can be gapped out to form two topological band gaps
(ideal for SHG) under an external magnetic field as shown in \cite{Jin_PRL17, Pan_NC17}. In fact,
recently, we have studied the topologically protected four-wave mixing process in a graphene
metasurface \cite{You_arxiv19}. \textit{iii) Mimicking time-reversal breaking and synthesizing
magnetic fields for photons} \cite{Rechtsman_Nature13, Hafezi_NP11, Hafezi_NP13}: using an array of
evanescently coupled helical waveguides, topological one-way edge states similar to that presented
in Fig.~\ref{fig3} can be created without the need of an external magnetic field (see, e.g.,
Fig.~2b in Ref.~\cite{Rechtsman_Nature13}). Alternatively, one can also create  one-way edge states
using synthetic magnetic fields rather than a real magnetic field in an array of coupled
optical-ring resonators (e.g., see Fig.~4 in Ref.~\cite{Hafezi_NP13}). Furthermore, additional
topological band gaps at higher frequencies can be readily created by considering waveguide modes
at corresponding frequencies. \textit{iv) Photonic quantum valley} \cite{Noh_PRL18, Shalaev_NN19,
He_NC19, Chen_AOM19} \textit{or spin} \cite{Yuri_PRL2019, Hu_PRL15, Barik_Science18, Peng_PRL19,
Noh_NP18, Mittal_PRL19, Shi_PNAS17} \textit{Hall (QVH or QSH) crystals with time-reversal
symmetry}: recent experiments on valley-Hall-like photonic systems have demonstrated the existence
of one \cite{Noh_PRL18, Shalaev_NN19, He_NC19} or two \cite{Chen_AOM19} (see Fig.~2b therein)
topological band gaps and corresponding edge modes within the gaps. As for the photonic QSH
systems, the experiments in \cite{Yuri_PRL2019, Barik_Science18, Peng_PRL19} have demonstrated the
existence of one topological band gap. Moreover, the proposal in \cite{Shi_PNAS17} (see Figs.~3 and
4 therein) using optically-passive elements to realize the optical version of the QSH insulator has
shown the existence of multiple topological band gaps.

\section{Conclusions}\label{concl}
In this work, we have demonstrated topology-protected nonlinear frequency conversion processes via
one-way edge modes of topological photonic crystals. Apart from the proof-of-concept
implementations, such as SHG and THG, we also showed that more complex behaviors, such as
slow-light effects and counter-propagating mode interaction, can also be realized within the setup.
A special aspect of nonlinear processes, i.e., the phase-matching condition, requires a new level
of control of the edge modes, which has not been discussed previously. This condition requires a
method to tune the dispersion of the edge modes, which we found can be conveniently achieved by
tailoring the geometry of the edge termination. Our work reveals that the coherence length
characterizing the nonlinear optical interactions considered in this study, which crucially
determines the efficiency of the topologically protected nonlinear frequency mixing processes, is
robust against structural defects. Our proposed setup provides a platform for studying additional
phenomena, e.g., when the frequency gap has large Chern number ($|C|\neq 0,1$ \cite{Skirlo_PRL04})
one can explore how to excite one of the several edge modes in the gap or how to couple edge modes
belonging to different gaps via the nonlinearity of the medium.

Importantly, nonlinear interactions of topological modes, such as sum- and difference-frequency
generation, high-harmonic generation, and four-wave-mixing, can be readily implemented within our
setup. Our work may also stimulate the search for other lattice geometries or setups where one can
optimize the gap properties for specific applications. For example, in the Chern number graded gap
phase diagrams of Fig.~\ref{fig2}e, apart from the gap of $C=1$, other gaps with $C=-1,-2,2,3$ are
typically narrow and appear at high frequencies, so designing setups where these gaps are wide and
are formed at low frequencies is particularly relevant from experimental point of view. Beyond the
experimental implementation of our model system at microwave frequencies using magnetic and
nonlinear materials, we also discussed several different possible implementations using diverse
experimental platforms available for topologic photonics. Last but not least, the concept of
topology-protected nonlinear frequency mixing is very general in that it applies not only to
photonics, but also to plasmonics \cite{Jin_PRL17, Pan_NC17, Kauranen_NP12, Panoiu_JO18}, phononics
\cite{Zhang_PRL15, Bojahr_PRL15, Ge_NSR18}, magnonics \cite{Chumak_NP15, Sebastian_PRL13,
Wang_PRAPP18} and exciton-polariton systems \cite{Karzig_PRX15, Klembt_Nature18, Warkentin_PRB18,
Knuppel_Nature19}, thus we expect that our work will have a broad impact.

\section*{Acknowledgements}
This work was supported by the European Research Council (ERC) (Grant no. ERC-2014-CoG- 648328). We
acknowledge the use of the UCL Legion High Performance Computing Facility (Legion@UCL) and
associated support services in the completion of this work.

\appendix
\begin{appendices}
\section{Computation of Chern numbers}
\label{appsec:1} The Chern number of the $n$th-band of the photonic crystal is defined by
Eq.~(\ref{Cn}) based on Refs.~\cite{Haldane_PRL08, Lu_NP14}. An efficient numerical algorithm to
compute it was introduced in Ref.~\cite{Fukui_JPSJ05}. In this method, from the eigenmode
$|{\mathbf{E}_{n}(\mathbf{k})}\rangle$ of the $n$th band, which can be accessed directly from the
eigenfrequency solver of COMSOL, one can define a  $U(1)$ link variable,
\begin{gather}
U_{\alpha}(\mathbf{k}_l)=\frac{{\left\langle {{{\mathbf{E}}_n}({{\mathbf{k}}_l})\left| {{{\mathbf{E}}_n}({{\mathbf{k}}_l} + {{\mathbf{e}}_\alpha })} \right.} \right\rangle }}{{\left| {\left\langle {{{\mathbf{E}}_n}({{\mathbf{k}}_l})\left| {{{\mathbf{E}}_n}({{\mathbf{k}}_l} + {{\mathbf{e}}_\alpha })} \right.} \right\rangle } \right|}},
\end{gather}
where $\mathbf{k}_l$ is a lattice point in the discretized Brillouin zone and $\mathbf{e}_{\alpha}$
is the lattice displacement in the direction $\alpha$ $(\alpha=1,2)$. Furthermore, a lattice field
strength can be defined by the link variable:
\begin{gather}
F_{12}(\mathbf{k}_l)= \ln\left[U_1(\mathbf{k}_l) U_2(\mathbf{k}_l+\mathbf{e}_{1})
U_1^{-1}(\mathbf{k}_l+\mathbf{e}_{2}) U_2^{-1}(\mathbf{k}_l)\right],
\end{gather}
where the lattice field strength is defined as the principal branch of the logarithm $-\pi
<F_{12}((\mathbf{k}_l)/i\leq\pi$. Finally, the Chern number of the $n$th band can be calculated
from the lattice field strength according to:
\begin{gather}
C_n=\frac{1}{2\pi i}\sum_l F_{12}(\mathbf{k}_l),
\end{gather}
where the sum is taken over all the lattice points in the discretized Brillouin zone. The $C_n$
defined above is manifestly gauge-invariant and strictly an integer for arbitrary lattice
parameters. This is because if we introduce a gauge potential
\begin{gather}
A_{\alpha}(\mathbf{k}_l)=\ln U_{\alpha}(\mathbf{k}_l), \quad -\pi
<A_{\alpha}(\mathbf{k}_l)/i\leq\pi,
\end{gather}
one can get
\begin{gather}
F_{12}(\mathbf{k}_l)=\Delta_1A_{2}(\mathbf{k}_l)-\Delta_2A_{1}(\mathbf{k}_l)+i2\pi
n_{12}(\mathbf{k}_l),
\end{gather}
where $\Delta_{\alpha}f(\mathbf{k}_l)=f(\mathbf{k}_l+\mathbf{e}_{\alpha})-f(\mathbf{k}_l)$ and $n_{12}(\mathbf{k}_l)$ is an integer valued field, chosen in such a way that $F_{12}(\mathbf{k}_l)/i$ takes a value within the principal branch. Consequently, we can get
\begin{gather}
C_n=\sum_l n_{12}(\mathbf{k}_l),
\end{gather}
which shows that $C_n$ is an integer. Certainly, this does not mean that any coarse discretization
of the first Brillouin zone will ensure a converged Chern number; nevertheless, asymptotic
convergence of the Chern number requires only a moderately dense discretization. In our
calculations of the data presented in Fig.~\ref{fig2}e, we find that a $20 \times 20$
discretization of the first Brillouin zone suffices.

\section{Coupled-mode theory for the third-harmonic generation}\label{appsec:2}
The coupled-mode equations for the third-harmonic generation can be derived in a way similar to
that for the second-harmonic generation presented in Section~\ref{CMT-SHG}. Here we outline the
main steps \cite{pmw10jstqe,lp16prb}. First, one can write down the expressions for the fundamental
($f$) and third ($t$) harmonic waves, $(\mathbf{E}_1^f, \mathbf{H}_1^f, \mathbf{E}_2^f,
\mathbf{H}_2^f, \mathbf{P}_{NL}^f)$ and $(\mathbf{E}_1^t, \mathbf{H}_1^t, \mathbf{E}_2^t,
\mathbf{H}_2^t, \mathbf{P}_{NL}^t)$, in the following Bloch forms,
\begin{subequations}
\begin{align}
\mathbf{E}_1^f(\mathbf{r},\omega_0)&=\frac{ \mathbf{e}_{f}(\mathbf{r},\omega_0)}{\sqrt{P_f}}e^{i k_{f}(\omega_0) x},\\
\mathbf{H}_1^f(\mathbf{r},\omega_0)&=\frac{\mathbf{h}_{f}(\mathbf{r},\omega_0)}{\sqrt{P_f}}e^{i k_{f} (\omega_0) x},\\
\mathbf{E}_1^t(\mathbf{r},\Omega_3)&=\frac{ \mathbf{e}_{t}(\mathbf{r},\Omega_3)}{\sqrt{P_t}}e^{i k_{t}(\Omega_3) x},\\
\mathbf{H}_1^t(\mathbf{r},\Omega_3)&=\frac{\mathbf{h}_{t}(\mathbf{r},\Omega_3)}{\sqrt{P_t}}e^{i
k_{t} (\Omega_3) x},
\end{align}
\end{subequations}
and
\begin{subequations}
\begin{align}
\mathbf{E}_2^f(\mathbf{r},\omega_0)&= A_f(x)\frac{\mathbf{e}_{f}(\mathbf{r},\omega_0)}{\sqrt{P_f}}e^{i k_{f}(\omega_0) x}, \\
\mathbf{H}_2^f(\mathbf{r},\omega_0)&=A_f(x)\frac{\mathbf{h}_{f}(\mathbf{r},\omega_0)}{\sqrt{P_f}}e^{i k_{f}(\omega_0) x},\\
\mathbf{E}_2^t(\mathbf{r},\Omega_3)&= A_t(x)\frac{ \mathbf{e}_{t}(\mathbf{r},\Omega_3)}{\sqrt{P_t}}e^{i k_{t}(\Omega_3)  x}, \\
\mathbf{H}_2^t(\mathbf{r},\Omega_3)&= A_t(x)\frac{
\mathbf{h}_{t}(\mathbf{r},\Omega_3)}{\sqrt{P_t}}e^{i k_{t}(\Omega_3)x},
\end{align}
\end{subequations}
where $\Omega_3=3\omega_0$ and $ \mathbf{P}_{NL}^f,  \mathbf{P}_{NL}^t$ can be derived from $\mathbf{P}_{NL}(\mathbf{r},t)=\hat{\chi}^{(3)}(\mathbf{r},t) \vdots \mathbf{E}(\mathbf{r},t) \mathbf{E}(\mathbf{r},t)\mathbf{E}(\mathbf{r},t)$ with $ \mathbf{E}(\mathbf{r},t)=\mathbf{E}_2^f(\mathbf{r},t)+\mathbf{E}_2^t(\mathbf{r},t)$  and $\hat{\chi}^{(3)}$ the third-order susceptibility tensor. After some simple algebra, the nonlinear polarizations at the fundamental and third-harmonic frequencies can be written as:
\begin{subequations}
\begin{align}
&\mathbf{P}_{NL}^{f}(\mathbf{r},\omega_0)=\frac{3{A_{f}^*}^{2}(x)A_t(x)}{P_f\sqrt{P_t}}\hat{\chi}^{(3)}\vdots \mathbf{e}_f^*(\mathbf{r},\omega_0)\mathbf{e}_f^*(\mathbf{r},\omega_0)\mathbf{e}_t(\mathbf{r},\Omega_3) e^{i(k_t-2k_f)x},\\
&\mathbf{P}_{NL}^{t}(\mathbf{r},\Omega_3)=\frac{A_f^{3}(x)}{P_f\sqrt{P_f}}\hat{\chi}^{(3)}\vdots\mathbf{e}_f(\mathbf{r},\omega_0)
\mathbf{e}_f(\mathbf{r},\omega_0)\mathbf{e}_f(\mathbf{r},\omega_0) e^{3i k_f x}.
\end{align}
\end{subequations}

If we now consider the fields $(\mathbf{E}_1^f, \mathbf{H}_1^f, \mathbf{E}_2^f, \mathbf{H}_2^f,
\mathbf{P}_{NL}^f)$, the l.h.s. of Eq.~\eqref{dt} gives:
\begin{equation}
\int_{-\infty}^{\infty} \nabla\cdot \mathbf{F} dy = \frac{3i\omega_0 {A_{f}^*}^{2} A_t e^{i\Delta k
x} }{P_f\sqrt{P_fP_t}}  \int_{-\infty}^{\infty} \mathbf{e}_f^{*}\cdot \hat{\chi}^{(3)}\vdots
\mathbf{e}_{f}^*\mathbf{e}_{f}^* \mathbf{e}_t dy, \label{f5}
\end{equation}
where $\Delta k=k_t(\Omega_3)-3k_f(\omega_0)$, and the r.h.s. of Eq.~\eqref{dt} leads to:
\begin{align}
\frac{\partial}{\partial x} &\int_{-\infty}^{\infty} \mathbf{F}\cdot\hat{\mathbf{x}} dy =
\frac{\partial}{\partial x} \int_{-\infty}^{\infty} \left(\mathbf{E}_1^{f*}\times
\mathbf{H}_2^f+\mathbf{E}_2^f\times \mathbf{H}_1^{f*}\right) \cdot\hat{\mathbf{x}}dy \nonumber
\\=& \frac{d A_f(x)}{d x}\frac{1}{P_f} \int_{-\infty}^{\infty}  \left(\mathbf{e}_f^*\times
\mathbf{h}_f+\mathbf{e}_f \times \mathbf{h}_f^*\right)  \cdot \hat{\mathbf{x}} dy  = 4\frac{d
A_f(x)}{dx}. \label{f6}
\end{align}

Comparing Eqs.~\eqref{f5} and \eqref{f6}, we obtain the coupled-mode equation for the
slowly-varying mode amplitude $A_f(x)$:
\begin{equation}
 \frac{d A_f(x)}{d x} =i\gamma^{(3)}_{f}(x) {A_{f}^*}^{2}(x) A_t(x) e^{i\Delta k x},
 \label{thg11}
\end{equation}
where the nonlinear coefficient at the fundamental frequency is
\begin{equation}
\gamma^{(3)}_{f}(x)=\frac{3\omega_0}{4P_f\sqrt{P_fP_t}} \int_{-\infty}^{\infty} \mathbf{e}_f^{*}
\cdot \hat{\chi}^{(3)}(\omega_0;-\omega_0, -\omega_0, \Omega_3)\vdots
\mathbf{e}_{f}^*\mathbf{e}_{f}^* \mathbf{e}_t dy. \label{thg12}
\end{equation}

Similarly, if we consider the fields $(\mathbf{E}_1^t, \mathbf{H}_1^t, \mathbf{E}_2^t,
\mathbf{H}_2^t, \mathbf{P}_{NL}^t)$, the l.h.s. of Eq.~\eqref{dt} gives:
\begin{equation}
\int_{-\infty}^{\infty} \nabla\cdot \mathbf{F} dy = \frac{i\Omega_3 A_{f}^{3} e^{i \Delta k
x}}{P_f\sqrt{P_f P_t}}  \int_{-\infty}^{\infty}  \mathbf{e}_t^{*}\cdot
\hat{\chi}^{(3)}(\Omega_3;\omega_0, \omega_0, \omega_0)\vdots  \mathbf{e}_{f}\mathbf{e}_{f}
\mathbf{e}_f dy, \label{f7}
\end{equation}
and the r.h.s. of Eq.~\eqref{dt} can be written as:
\begin{align}
\frac{\partial}{\partial x} &\int_{-\infty}^{\infty} \mathbf{F}\cdot\hat{\mathbf{x}} dy =
\frac{\partial}{\partial x} \int_{-\infty}^{\infty} \left(\mathbf{E}_1^{t*}\times
\mathbf{H}_2^t+\mathbf{E}_2^t\times \mathbf{H}_1^{t*} \right) \cdot\hat{\mathbf{x}}dy \nonumber
\\=& \frac{d A_t(x)}{d x}\frac{1}{P_t} \int_{-\infty}^{\infty}\left(\mathbf{e}_t^*\times
\mathbf{h}_t + \mathbf{e}_t \times \mathbf{h}_t^*\right)  \cdot \hat{\mathbf{x}} dy  = 4\frac{d
A_t(x)}{dx}. \label{f8}
\end{align}

Finally, from Eqs.~\eqref{f7} and \eqref{f8}, we get the coupled-mode equation for the
slowly-varying amplitude $A_t(x)$ as:
\begin{equation}
 \frac{d A_t(x)}{d x} =i\gamma^{(3)}_t(x)A_{f}^{3}(x)e^{-i\Delta k x},
 \label{thg21}
\end{equation}
where the nonlinear coefficient at the third-harmonic frequency is:
\begin{equation}
 \gamma^{(3)}_{t}(x)=\frac{\Omega_3}{4P_f\sqrt{P_fP_t}} \int_{-\infty}^{\infty} \mathbf{e}_t^{*}\cdot \hat{\chi}^{(3)}(\Omega_3;\omega_0, \omega_0, \omega_0)\vdots \mathbf{e}_{f}\mathbf{e}_{f}\mathbf{e}_fdy.
 \label{thg22}
\end{equation}

We can also express the nonlinear coefficients in terms of the group indices and the effective
third-order susceptibilities, $\chi_{\mathrm{eff},f/t}^{(3)}(x)$, of the two interacting modes:
\begin{subequations}\label{nonlcoeff3}
\begin{align}
\gamma_{f}^{(3)}(x)& =\frac{12\omega_0}{a\epsilon_{0}^2v_{g,f}\sqrt{v_{g,f}v_{g,t}}}\chi_{\text{eff},f}^{(3)}(x), \\
\gamma_{t}^{(3)}(x)&=\frac{4\Omega_3}{a\epsilon_{0}^2v_{g,f}\sqrt{v_{g,f}v_{g,t}}}\chi_{\text{eff},t}^{(3)}(x),
\end{align}
\end{subequations}
where
\begin{widetext}
\begin{subequations}\label{effchi3}
\begin{align}
\chi_{\mathrm{eff},f}^{(3)}(x)&=\frac{\displaystyle a^{3}\int_{-\infty}^{\infty}
\mathbf{e}_f^{*} \cdot \hat{\chi}^{(3)}(\omega_0;-\omega_0, -\omega_0, \Omega_3)\vdots \mathbf{e}_{f}^*\mathbf{e}_{f}^* \mathbf{e}_t dy}{\displaystyle \left[ \int_{A_{cell}}\left(\mathbf{e}_{f}^{*}\cdot\hat{\epsilon}_{r}\mathbf{e}_{f}+Z_{0}^{2}\mathbf{h}_{f}^{*}\cdot\hat{\mu}_{r}\mathbf{h}_{f}\right)dA \right]^{3/2}
{\left[\int_{A_{cell}}\left(\mathbf{e}_{t}^{*}\cdot\hat{\epsilon}_{r}\mathbf{e}_{t}+Z_{0}^{2}\mathbf{h}_{t}^{*}\cdot\hat{\mu}_{r}\mathbf{h}_{t}\right)dA\right]}^{1/2}},\\
\chi_{\mathrm{eff},t}^{(3)}(x)&=\frac{\displaystyle a^{3}\int_{-\infty}^{\infty}
\mathbf{e}_t^{*}\cdot \hat{\chi}^{(3)}(\Omega_3;\omega_0, \omega_0, \omega_0)\vdots \mathbf{e}_{f}\mathbf{e}_{f}\mathbf{e}_fdy}{\displaystyle \left[ \int_{A_{cell}}\left(\mathbf{e}_{f}^{*}\cdot\hat{\epsilon}_{r}\mathbf{e}_{f}+Z_{0}^{2}\mathbf{h}_{f}^{*}\cdot\hat{\mu}_{r}\mathbf{h}_{f}\right)dA\right]^{3/2}
{\left[\int_{A_{cell}}\left(\mathbf{e}_{t}^{*}\cdot\hat{\epsilon}_{r}\mathbf{e}_{t}+Z_{0}^{2}\mathbf{h}_{t}^{*}\cdot\hat{\mu}_{r}\mathbf{h}_{t}\right)dA\right]}^{1/2}}.
\end{align}
\end{subequations}
\end{widetext}

Introducing the averaged physical quantities,
\begin{subequations}
\begin{align}
\overline{\gamma}_{f,t}^{(3)}&=\frac{1}{a} \int_{x_0}^{x_0+a}\gamma_{f,t}^{(3)}(x) dx, \\
\overline{\gamma}_{\text{eff},f/t}^{(3)}&=\frac{1}{a}
\int_{x_0}^{x_0+a}\gamma_{\text{eff},f/t}^{(3)}(x)dx,
\end{align}
\end{subequations}
and then averaging Eqs.~\eqref{thg11} and \eqref{thg22} over one lattice constant, we obtain the
coupled-mode equations governing the nonlinear dynamics of the interacting modes:
\begin{subequations}\label{CMTaverage3}
\begin{align}
\frac{d \overline{A}_f(x)}{d x}&=i\overline{\gamma}^{(3)}_f {\overline{A}_{f}^*}^{2}(x)\overline{A}_t(x) e^{i\Delta k x},   \\
\frac{d \overline{A}_t(x)}{d x}&=i\overline{\gamma}^{(3)}_t \overline{A}_{f}^{3}(x)e^{-i\Delta k
x},
\end{align}
\end{subequations}
where $\overline{A}_f(x)$ and $\overline{A}_t(x)$ are the averaged mode amplitudes at the
fundamental and third-harmonic frequencies, respectively.
\begin{figure}[!b]
\centering
\includegraphics[width=\columnwidth]{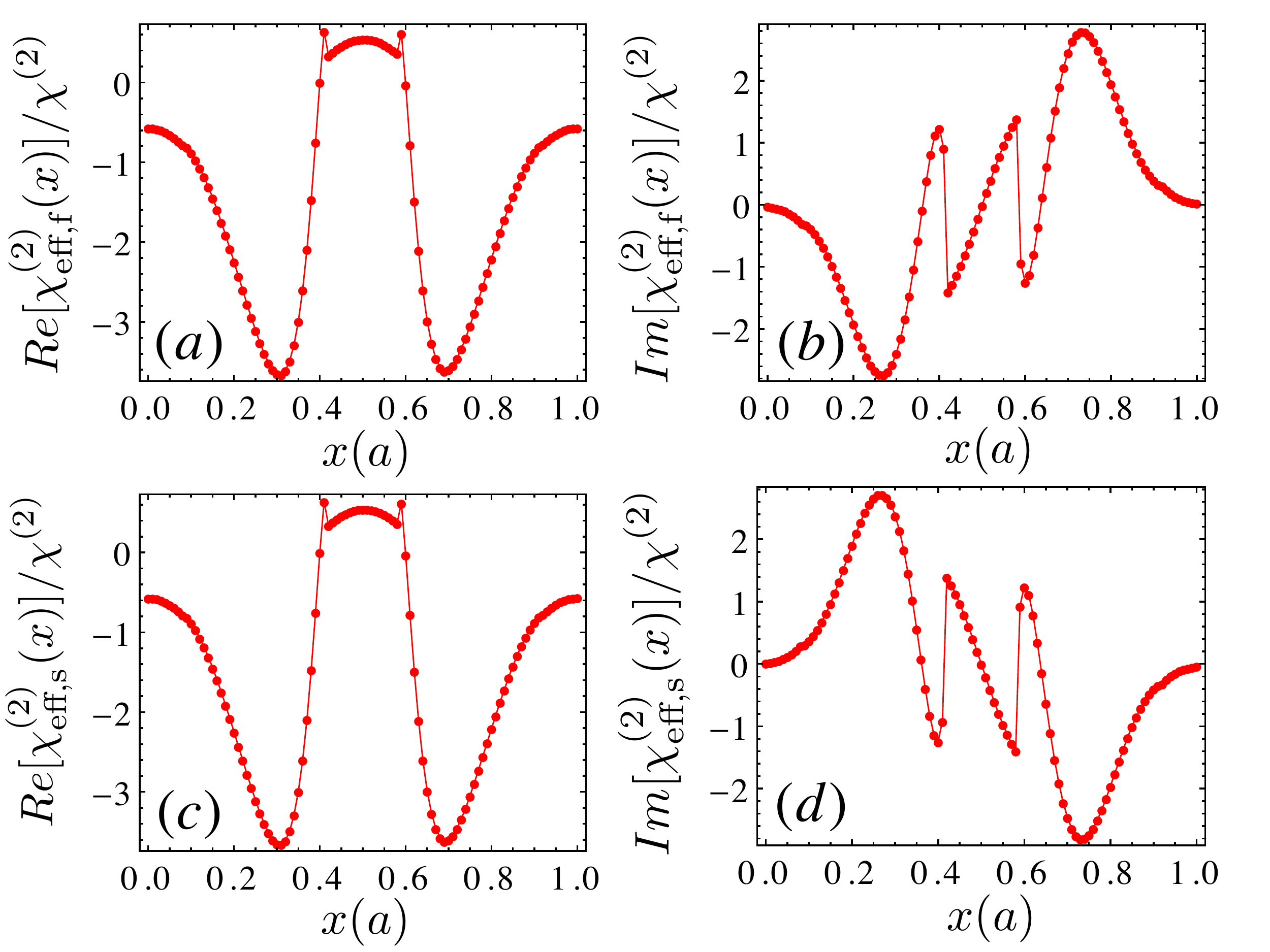}
\caption{Real (a, c) and imaginary (b, d) parts  of $\chi^{(2)}_{\textrm{eff},f}(x)$ and
$\chi^{(2)}_{\textrm{eff},s}(x)$ of Eq. (\ref{effchi2}) for SHG in one unit cell. While the real
parts of $\chi^{(2)}_{\textrm{eff},f}(x)$ and $\chi^{(2)}_{\textrm{eff},s}(x)$ are even functions,
with respect to the center of the unit cell, the imaginary parts of
$\chi^{(2)}_{\textrm{eff},f}(x)$ and $\chi^{(2)}_{\textrm{eff},s}(x)$ are odd
functions.}\label{figs1}
\end{figure}

\section{Comparison between rigorous numerical simulations and the coupled-mode theory}\label{appsec:3}
In this Appendix, we present a computational analysis that illustrates how the coupled-mode theory
derived above can be used to explain the full-wave dynamics obtained using COMSOL.
\begin{figure}[!b]
\centering
\includegraphics[width=\columnwidth]{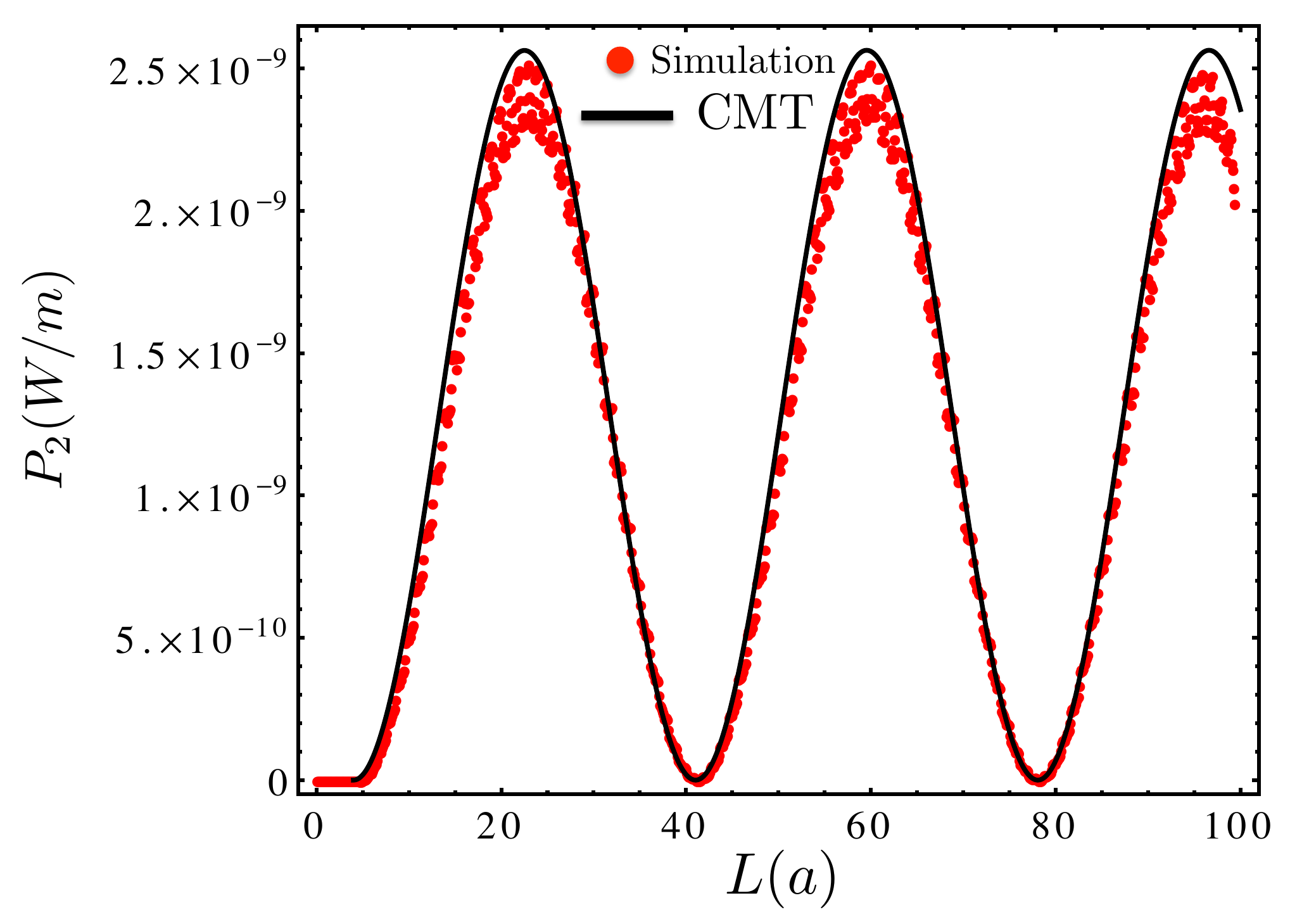}
\caption{The evolution of the power of the second-harmonic wave as a function of propagation
distance. The red dots are from rigorous full-wave numerical simulations while the black curve is
obtained from solving the coupled-mode equations expressed as
Eqs.~\eqref{CMTaverage2}.}\label{figs2}
\end{figure}

\subsection{Second-harmonic generation described by the coupled-mode theory}
\label{appsec:31} The key quantities that characterize the coupled-mode theory describing the
second-harmonic generation are $\chi^{(2)}_{\textrm{eff},f}(x)$ and
$\chi^{(2)}_{\textrm{eff},s}(x)$, as defined by Eqs.~\eqref{effchi2f} and \eqref{effchi2s},
respectively. As we consider photonic crystals that are periodic in space,
$\chi^{(2)}_{\textrm{eff},f}(x)$ and $\chi^{(2)}_{\textrm{eff},s}(x)$ are periodic functions of
$x$, and we only need to show their $x$-dependence in one unit cell. Thus, in Fig.~\ref{figs1} we
depict the $x$-dependence of $\chi^{(2)}_{\textrm{eff},f}(x)$ and $\chi^{(2)}_{\textrm{eff},s}(x)$
in one unit cell. As one can see, while the real parts of $\chi^{(2)}_{\textrm{eff},f}(x)$ and
$\chi^{(2)}_{\textrm{eff},s}(x)$ are even functions, with respect to the center of the unit cell,
the imaginary parts of $\chi^{(2)}_{\textrm{eff},f}(x)$ and $\chi^{(2)}_{\textrm{eff},s}(x)$ are
odd functions. In fact, it can be easily demonstrated that
$\chi^{(2)}_{\textrm{eff},f}(x)={\chi^{(2)}_{\textrm{eff},s}}^{*}(x)$. Therefore, the averages of
$\chi^{(2)}_{\textrm{eff},f}(x)$ and $\chi^{(2)}_{\textrm{eff},s}(x)$ in one unit cell are real
numbers and are equal to each other.

After one computes $\overline{\gamma}^{(2)}_{f,s}$ and $\overline{\chi}^{(2)}_{\mathrm{eff},f/s}$
by averaging $\gamma^{(2)}_{f,s}(x)$ and $\chi^{(2)}_{\mathrm{eff},f/s}(x)$, respectively, in one
unit cell, one can straightforwardly solve the coupled-mode equations expressed as
Eqs.~\eqref{CMTaverage2}. Thus, we present in Fig.~\ref{figs2} the evolution of the generated
second-harmonic wave as a function of propagation distance. It can be seen in this figure that
there is a good agreement between the coupled-mode theory and rigorous numerical simulations both
in regard of the oscillation period and the amplitude of the power of the second-harmonic wave.

\subsection{Third-harmonic generation described by the coupled-mode theory}
\label{appsec:32} 
\begin{figure}[!b]
\centering
\includegraphics[width=\columnwidth]{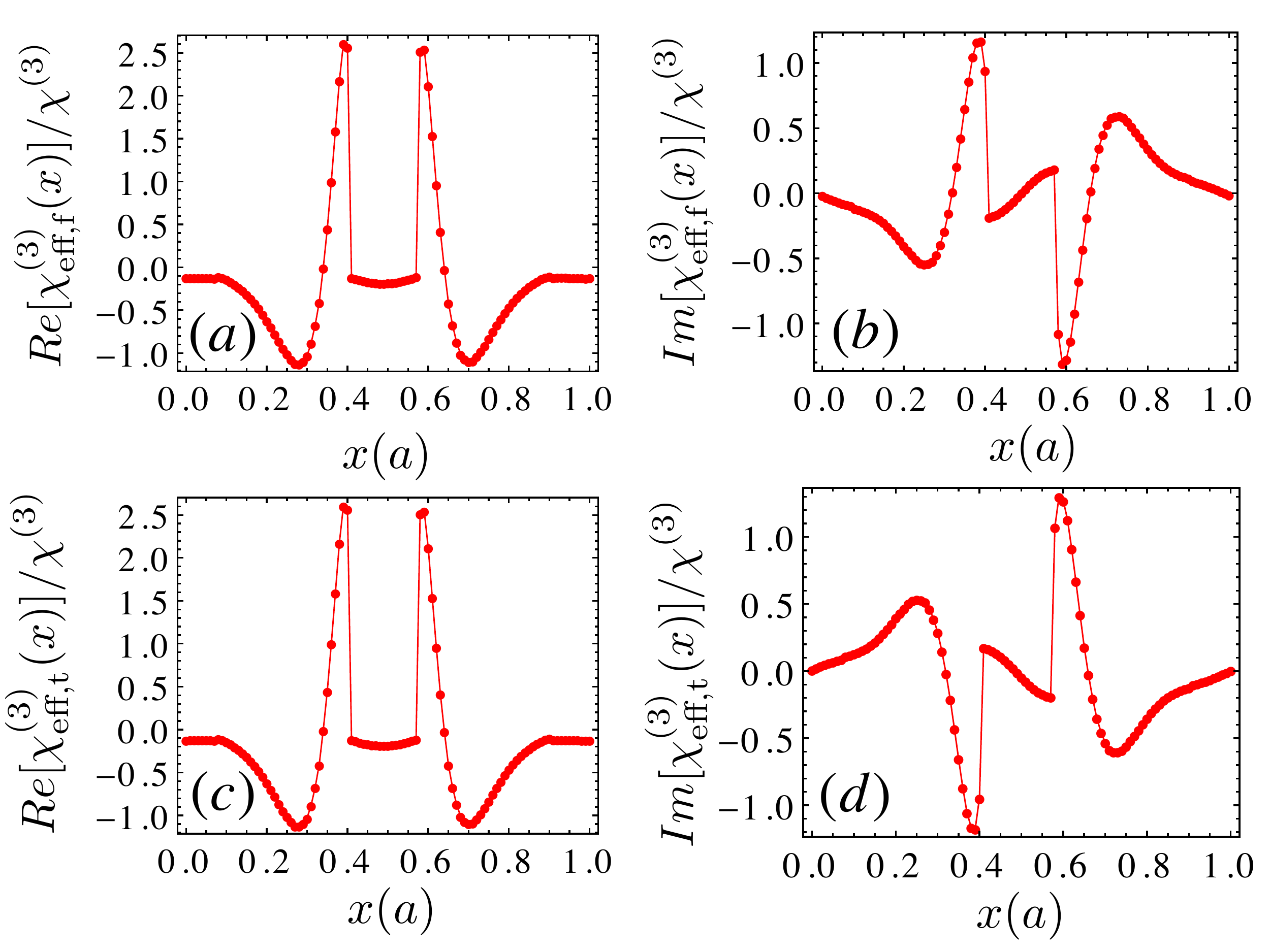}
\caption{Real (a, c) and imaginary (b, d) parts of $\chi^{(3)}_{\textrm{eff},f}(x)$ and
$\chi^{(3)}_{\textrm{eff},t}(x)$ of Eqs.~\eqref{effchi3} for THG in one unit cell. Similar to the
case of SHG, while the real parts of $\chi^{(3)}_{\textrm{eff},f}(x)$ and
$\chi^{(3)}_{\textrm{eff},t}(x)$ are even functions, the imaginary parts of
$\chi^{(3)}_{\textrm{eff},f}(x)$ and $\chi^{(3)}_{\textrm{eff},t}(x)$ are odd
functions.}\label{figs3}
\end{figure}

Similar to the case of SHG, the key quantities that characterize the coupled-mode
theory of third-harmonic generation are $\chi^{(3)}_{\textrm{eff},f}(x)$ and
$\chi^{(3)}_{\textrm{eff},t}(x)$, as defined by Eqs.~\eqref{effchi3}. Both these physical
quantities are periodic functions of the $x$-coordinate, their $x$-dependence being presented in
Fig.~\ref{figs3}. Also similar to the case of SHG, while the real parts of
$\chi^{(3)}_{\textrm{eff},f}(x)$ and $\chi^{(3)}_{\textrm{eff},t}(x)$ are even functions, the
imaginary parts of $\chi^{(3)}_{\textrm{eff},f}(x)$ and $\chi^{(3)}_{\textrm{eff},t}(x)$ are odd
functions. Therefore, the averages of $\chi^{(3)}_{\textrm{eff},f}(x)$ and
$\chi^{(3)}_{\textrm{eff},t}(x)$ in one unit cell are real numbers and are equal to each other.

\begin{figure}[!t]
\centering
\includegraphics[width=\columnwidth]{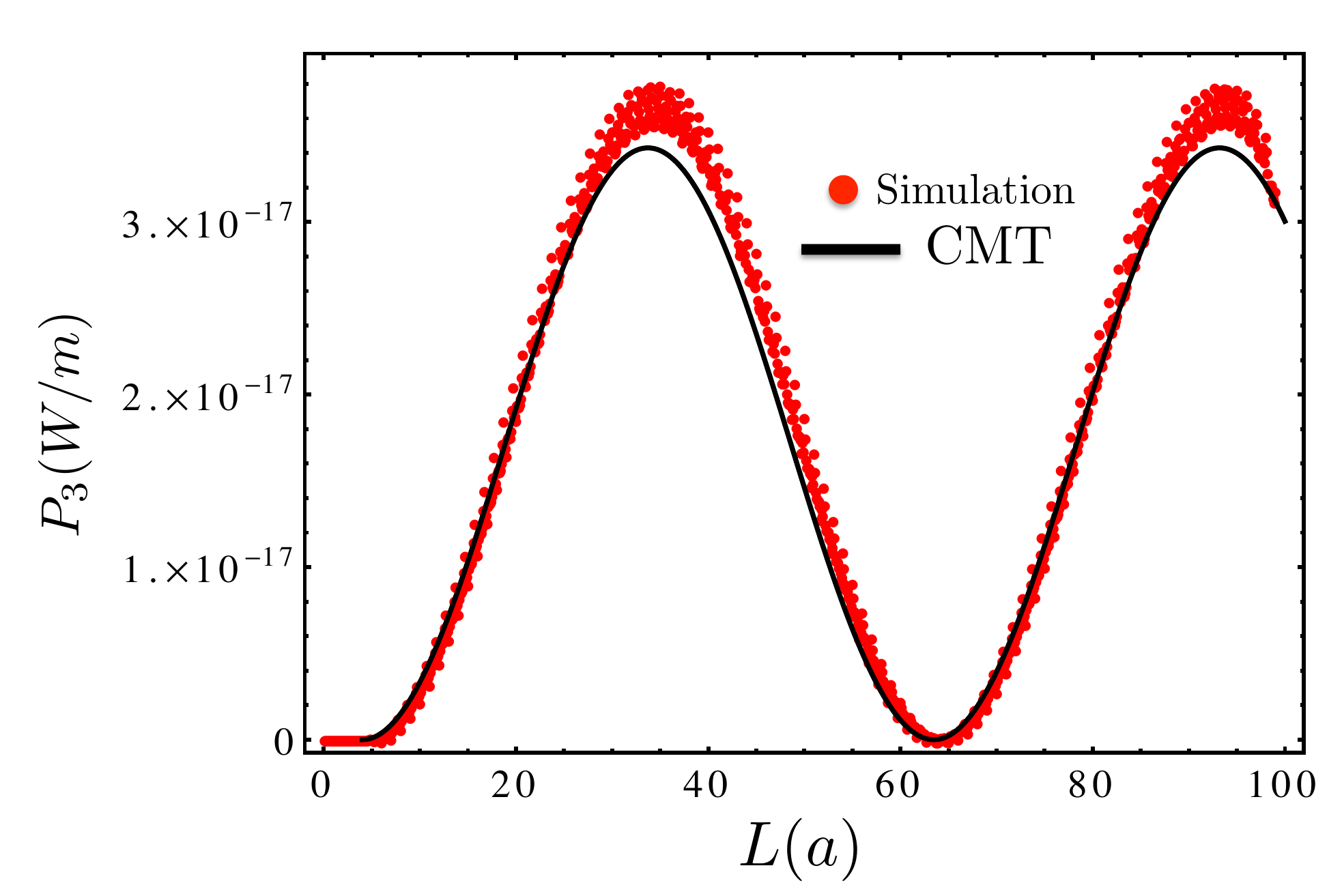}
\caption{The evolution of the power of the third-harmonic wave vs. propagation distance. The red
dots are from numerical simulations while the black curve is from solving
Eqs.~\eqref{CMTaverage3}.}\label{figs4}
\end{figure}

We have solved Eqs.~\eqref{CMTaverage3} using $\overline{\gamma}^{(3)}_{f,t}$ obtained by averaging
$\gamma^{(3)}_{f,t}(x)$ in one unit cell and present the results in Fig.~\ref{figs4}. We again find
a good agreement between the coupled-mode theory and rigorous numerical simulations regarding both
the oscillation period and amplitude of the power of the third-harmonic wave.

\begin{figure}[!b]
\centering
\includegraphics[width=\columnwidth]{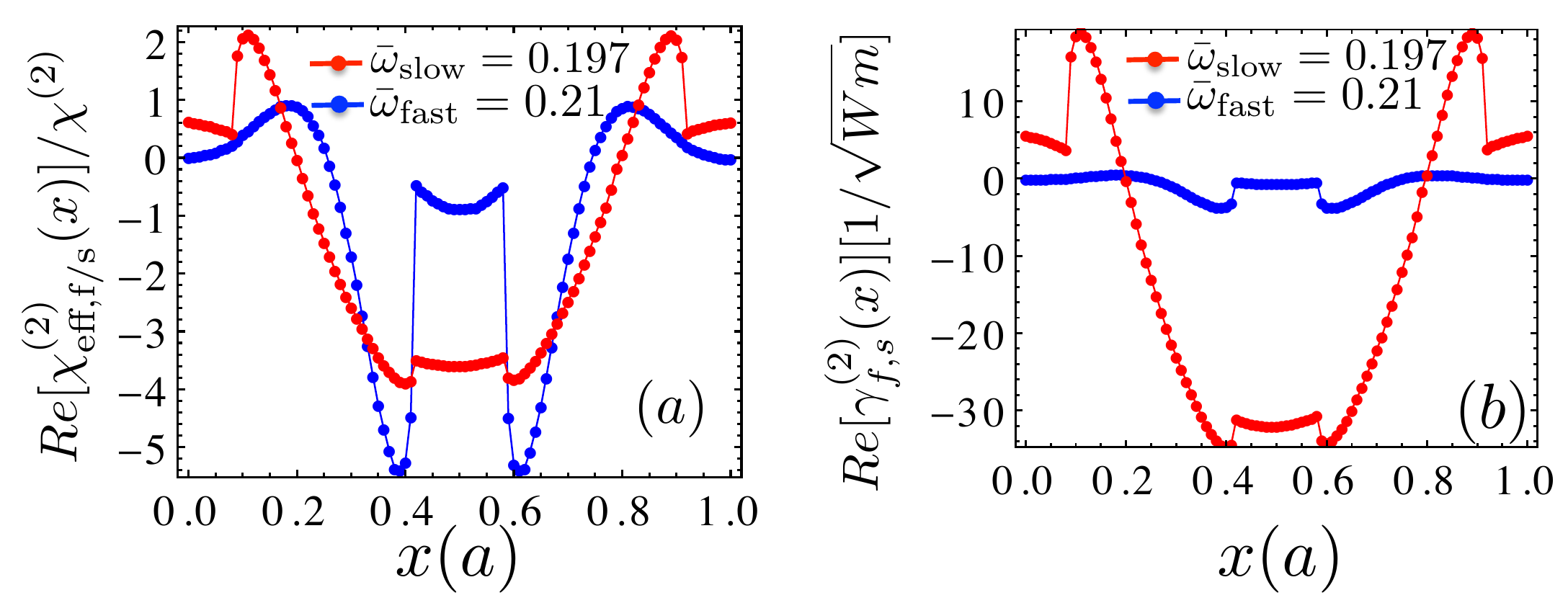}
\caption{ (a) $\chi^{(2)}_{\textrm{eff},f/s}(x)$ defined by Eqs.~\eqref{effchi2f}/\eqref{effchi2s}
for SHG in one unit cell at two different frequencies, $\bar{\omega}_{\textrm{slow}}=0.197$ and
$\bar{\omega}_{\textrm{fast}}=0.21$, where the  group indices of the fundamental wave at the two
frequencies are $n_g^f(\bar{\omega}_{\textrm{slow}}) \simeq174$ and
$n_g^f(\bar{\omega}_{\textrm{fast}})\simeq 12$ [$ n_g^s(\bar{\omega}_{\textrm{slow}}) \simeq 5$ and
$n_g^s(\bar{\omega}_{\textrm{fast}})\simeq 4$]. (b) The nonlinear coefficient
$\gamma_{f/s}^{(2)}(x)$ of Eqs.~\eqref{shg12}/\eqref{shg22} in one unit cell at the two frequencies
$\bar{\omega}_{\textrm{slow}}$ and $\bar{\omega}_{\textrm{fast}}$. One can see the enhancement of
the nonlinear coefficient due to the slow-light effect
[$n_g^f(\bar{\omega}_{\textrm{slow}})/n_g^f(\bar{\omega}_{\textrm{fast}})\simeq15$, whereas $
\bar{\gamma}^{(2)}_{f,s}(\bar{\omega}_{\textrm{slow}})/\bar{\gamma}^{(2)}_{f,s}(\bar{\omega}_{\textrm{fast}})
\simeq18$ according to Eq.~\eqref{average}].}\label{figs5}
\end{figure}
\subsection{Coupled-mode theory of slow-light nonlinearity enhancement}
\label{appsec:33} The group index $n_g$ of the edge modes in a generic case, e.g., Fig.~\ref{fig4},
is limited to 5$\sim$7. However, we have shown in the main text that the shape of the dispersion
curve of the one-way edge modes can be tailored, leading to much larger $n_g$, as per
Fig.~\ref{fig5}. As it is well known, this will enhance the efficiency of the nonlinear process. We
plot in Fig.~\ref{figs5}a the function $\chi^{(2)}_{\textrm{eff},f/s}(x)$ at two different
frequencies, $\bar{\omega}_{\textrm{slow}}=0.197$ and $\bar{\omega}_{\textrm{fast}}=0.21$ with
$n_g^f(\bar{\omega}_{\textrm{slow}}) \simeq174$ and $n_g^f(\bar{\omega}_{\textrm{fast}})\simeq 12$
[$ n_g^s(\bar{\omega}_{\textrm{slow}}) \simeq 5$ and $n_g^s(\bar{\omega}_{\textrm{fast}})\simeq 4$
have similar values at these two frequencies]. While the amplitudes of
$\chi^{(2)}_{\textrm{eff},f/s}(x)$ at these two frequencies are comparable, from Fig.~\ref{figs5}b,
which shows the nonlinear coefficient $\gamma_{f/s}^{(2)}(x)$ of Eqs.~\eqref{shg12}/\eqref{shg22},
one can see that $\gamma_{f/s}^{(2)}(x)$ at $\bar{\omega}_{\textrm{slow}}$ is significantly
enhanced as compared to $\gamma_{f/s}^{(2)}(x)$ at $\bar{\omega}_{\textrm{fast}}$. The ratio of the
averaged nonlinear coefficient in one unit cell according to Eq.~\eqref{average} and
Fig.~\ref{fig5}b is
$\bar{\gamma}^{(2)}_{f,s}(\bar{\omega}_{\textrm{slow}})/\bar{\gamma}^{(2)}_{f,s}(\bar{\omega}_{\textrm{fast}})
\simeq18$, which roughly agrees with
$n_g^f(\bar{\omega}_{\textrm{slow}})/n_g^f(\bar{\omega}_{\textrm{fast}})\simeq15$. This enhancement
of the nonlinear coefficient leads to the enhancement of the efficiency of the SHG, see, e.g.,
Figs.~\ref{fig5}c and \ref{fig5}d.

\end{appendices}

\end{document}